\documentclass[11pt]{article}
\usepackage{jheppubmod}
\pdfoutput=1
\usepackage{psfrag}
\usepackage{comment}
\usepackage{array}
\usepackage{amssymb}
\usepackage{amsmath}
\usepackage{amsthm}
\usepackage{graphicx}
\usepackage{caption}
\usepackage{float}

\title{Anisotropic inflation in Gauss-Bonnet gravity}
\author[a]{Sayantani Lahiri}

\affiliation[a]{ZARM, University of Bremen, \\
Am Falltrum,
28359 Bremen, Germany}
\emailAdd{sayantani.lahiri@gmail.com}

\bigskip

\abstract{ We study anisotropic inflation with Gauss-Bonnet correction in presence of a massless vector field. 
In this scenario, exact anisotropic power-law inflation is realized when the inflaton potential, gauge coupling function and
the Gauss-Bonnet coupling are exponential functions. 
We show that anisotropy becomes proportional
to two slow-roll parameters of the theory and hence gets enhanced in presence of quadratic curvature corrections. 
The stability analysis reveals that anisotropic power-law solutions remain stable 
over a substantially large parameter region.}

\begin{document}
\maketitle

\section{Introduction}
The inflationary paradigm \cite{infl1} 
in modern cosmology is immensely successful in resolving a number of limitations
of hot big bang model.
These drawbacks involve horizon problem, flatness problem, 
magnetic monopole problem, interpretations concerning sources of temperature fluctuations in CMB and 
large scale structure formation in universe \cite{Mukh}.
As a result, the inflationary phase, characterized by a period of accelerated expansion 
is considered as an integral part of
Standard model of cosmology.
In addition, inflation also predicts a nearly scale invariant primordial power spectrum \cite{Lyth}, almost adiabatic and 
Gaussian perturbations \cite{Juan,Planck1,Planck2} on CMB as well as statistical isotropy thus 
confirming the validity of cosmological principle at large scales.\\
Recent cosmological observations of WMAP and Planck \cite{WMAP, Planck} corroborate statistical isotropy,
homogeneity and flatness of the universe however not all observational data comply with predictions, 
for example, WMAP data \cite{WMAP2}
indicate scale dependence of power-spectrum and deviations 
from Gaussian nature of primordial perturbations. 
These deviations being quite small, are thought of as corrections to zeroth order predictions  
of inflation.
Nevertheless, zeroth order predictions match well with observational data. 
However, due to availability of high accuracy CMB data and other observational signatures, it is legitimate
to look beyond zeroth order results and concentrate on precision cosmology 
which involves studying fine structures of primordial fluctuations
such as spectral tilt, parity violation, low power  of
quadrupole \cite{Hin}.\\
Presently, it is known that universe does not exhibit exact de-Sitter nature (\cite{Jiro} and references therein)
and violations of both temporal and spatial de-Sitter symmetries occur.
In fact, violation of spatial de-Sitter symmetry leads to 
statistical anisotropy which has been confirmed observationally by Planck collaboration \cite{Planck2, Planck3}.
Motivated by these observations, several theoretical propositions came up 
to address and realize anisotropy generated during the inflationary phase \cite{Kar, Dim, Val, Yokoyama, Yamamato,Sheikh1}.
In \cite{Watanabe} stable anisotropic inflation solutions are obtained as a consequence of 
the back reaction of a vector field in the context of supergravity.\\ [1mm]
The energy scale at which inflation is believed to occur suggests that 
quantum corrections of gravity must be taken into account. 
This in turn demands formulation of a more fundamental theory by incorporating the gravity quantization in order to
describe Planck scale physics. 
 To date, superstring theory provides the most consistent set-up for quantum gravity 
 involving extra dimensions \cite{Green}. 
However, on the four dimensional world and in the low-energy limit, the reminiscence of the higher dimensional theories 
 like superstring theory appear as corrections terms of higher orders in the 
 curvature to the lowest order gravitational action. 
The simplest such correction in four dimensions is the Gauss-Bonnet term in the context of heterotic 
superstring effective theory\cite{Zwibach,Gross}.
Furthermore, the Gauss-Bonnet term is the first order correction term of the general extension
of the Einstein's gravity commonly known as Lovelock theory \cite{Lovelock} which gives rise to ghost-free theories.\\
 The Gauss-Bonnet term is topologically invariant in four dimensions and
does not modify gravitational equations of motion unless it is
non-minimally coupled to a scalar field. 
Therefore, when the Gauss-Bonnet term is non-minimally coupled to an additional scalar field, it 
influences dynamical equations in four dimensions.
So, studying anisotropic effects during the inflationary phase in presence of Gauss-Bonnet term
is therefore well-founded. 
In \cite{Watanabe}, anisotropic inflationary solutions are obtained with the help of a massless vector field
whose kinetic term is coupled to the inflaton field.   
We shall follow this approach here. 
We mention that isotropic inflation with Gauss-Bonnet corrections has been extensively studied in past
years \cite{Satoh1, Satoh2,Schwarz1, Koh, Kanti}.\\
Therefore, in the light of precision cosmology,
in the present work,  we investigate anisotropy effects 
during inflation in presence of Gauss-Bonnet gravity by coupling the inflaton field non-minimally
to the Gauss-Bonnet term.
The cosmological anisotropy is generated with the help of vector field which is also coupled 
to the inflaton.
The back-reaction of the vector field contributes non-trivially to the inflation dynamics and
exact power-law anisotropic solutions have been obtained when functional forms of the inflaton potential, 
gauge coupling function and the scalar-Gauss-Bonnet
coupling have exponential dependences on the inflaton field. 
Although small in magnitude, these effects do not get diluted during slow-roll regime. 
The anisotropic solutions so obtained have been found to be stable. \\
This outline of this paper is as follows. 
In section 2, we have showed how and under what conditions the anisotropic inflation can be realized in presence 
of Gauss-Bonnet term by taking the effect of a vector field and its back-reaction.
We have derived a general relation where anisotropy is found to be proportional to slow-roll parameters of
the theory. 
After constructing a general set-up for the anisotropic inflation,
in section 3 we obtain exact anisotropic power-law solutions in the Gauss-Bonnet gravity scenario.
We have estimated the anisotropy measure and compared with the results of 
non-Gauss-Bonnet case \cite{Kanno}.
In section 4, we have performed the stability analysis of obtained 
anisotropic power-law solutions after determining fixed phase points. 
Finally, we accumulate all results in the last section.

\section{Anisotropic inflationary solutions with Gauss-Bonnet gravity}
In order to generate anisotropic effects during inflation with the help of a vector field in presence of
Gauss-Bonnet correction term, we consider a massless vector field whose 
kinetic term is coupled to the inflaton field $\phi$
through the gauge coupling function $f(\phi)^2$. 
In this sense, $f(\phi)^2$ may also be called the gauge kinetic function.
Since we want to investigate the contribution of the Gauss-Bonnet term, it is therefore
non-minimally coupled to the inflaton field through the coupling function $\xi(\phi)$. 
With these considerations, the four-dimensional action is given by \\[1mm]
\begin{equation}
 S = \int d^{4}x \,\sqrt{-g} \left[\frac{R}{2 \kappa^2} -\frac{1}{2} \nabla _{\mu} \phi \nabla ^{\mu} \phi
 +\frac{1}{8}\xi(\phi) R^2_{GB}- V(\phi)-\frac{1}{4}f^2(\phi)F_{\mu \nu}F^{\mu \nu}\right],   \label{action}
\end{equation}
where the Gauss-Bonnet term $R^2_{GB}$ is 
\begin{equation}
R^2_{GB} = R_{\mu \nu \rho \beta}R^{\mu \nu \rho \beta}
-4 R_{\mu \nu}R^{\mu \nu}+R^2.
\end{equation}
Here, $\kappa^2$ is the four-dimensional gravitational constant, $V(\phi)$ is the inflaton potential and 
$F_{\mu \nu}=\partial_{\mu}A_{\nu}-\partial_{\nu}A_{\mu}$ is the electromagnetic field tensor.
By varying the action (\ref{action}) with respect to $g_{\mu \nu}$, the equation of motion is given by
\begin{equation}
\begin{array}{rcl}
 G_{\mu \nu} +\kappa^2 P_{\mu \alpha \nu \beta}\nabla^{\alpha}\nabla^{\beta}\xi & =&
 \kappa^2 \left[ \nabla_{\mu} \phi \nabla_{\nu} \phi -
 \frac{1}{2} g_{\mu \nu} 
 \nabla^{\beta}\phi \nabla_{\beta} \phi 
 -  V(\phi)g_{\mu \nu}\right]\\[2mm]
 &&  +  \kappa^2 f^2(\phi) \left( F_{\mu \alpha} F_{\nu \beta} g^{\alpha \beta}-
 \frac{1}{4} F_{\alpha \beta} F^{\alpha \beta} g_{\mu \nu} \right),  \label{grav_eqn}
 \end{array}
\end{equation}
where $G_{\mu \nu}$ is the Einstein's tensor and the Gauuss-Bonnet part in the equation of motion is given by
\begin{equation}
\begin{array}{rcl}
 P_{\mu \alpha \nu \beta}\nabla^{\alpha}\nabla^{\beta}\xi & = & R_{\mu \alpha \nu \beta}\nabla^{\alpha}\nabla^{\beta}\xi
 -\Box \xi R_{\mu \nu} \\[2mm]
 & +& \left(\nabla_{\mu}\nabla^{\alpha}\xi R_{\alpha \nu}+\nabla^{\beta} \nabla_{\nu} \xi R_{\mu \beta}\right)
 -\displaystyle\frac{1}{2}\nabla_{\mu}\nabla_{\nu}\xi \\[2mm]
 & - & \displaystyle\frac{1}{2}\left(2 \nabla^{\alpha}\nabla^{\beta}\xi R_{\alpha \beta}
 -R \Box \xi \right)g_{\mu \nu}.                        \label{GBform}
\end{array}
\end{equation}
The equation of motion of the inflaton field is given by
\begin{equation}
 \Box \phi +\frac{1}{8}\xi'(\phi)R^2_{GB}- V'(\phi)
 -\frac{1}{2}f'(\phi)f(\phi)F_{\mu \nu}F^{\mu \nu} = 0,     \label{box_scalar}
\end{equation}
where $'$ denotes derivative with respect to $\phi$.\\
It is to be noted that if the Gauss-Bonnet coupling $\xi(\phi)$ is constant, then $\nabla^{\alpha}\nabla^{\beta}\xi=0$
which implies that Gauss-Bonnet term adds up nothing to the equation of motion.
We mention here that isotropic inflation corresponds to the situation
when the gauge field $A_{\mu}$ identically vanishes to zero.
Without loss of generality, the $x-$axis is taken in the direction of the vector and
for simplicity the direction of the vector field is considered to be time independent.
Thus, in general the spatial isotropy is broken but due to our choice of the direction of the vector field 
the rotational symmetry exists in the $y-z$ plane.
With the gauge choice such that
the temporal component of electromagnetic potential satisfies $A_{0}=0$,
we can express $A_{\mu}=(0,v(t),0,0)$ from which the electromagnetic
field tensor $F_{\mu \nu}$ is constructed.  
We also assume that the inflaton field is a time-varying quantity i,e. $\phi=\phi(t)$.
Then, in order to find anisotropic scaling solutions in presence of Gauss-Bonnet term, we
consider the following Bianchi-I metric \cite{Watanabe} \\[1mm]
\begin{equation}
 ds^2 = -dt^2 +e^{2\alpha(t)}\left[ e^{-4\sigma(t)}dx^2 + e^{2\sigma(t)}(dy^2 + dz^2) \right],  \label{metric-aniso}
\end{equation}
where $t$ is the cosmic time. 
The isotropic scale factor is given by $e^{\alpha}$ and $\sigma$ specifies deviation from isotropy.
The equation of motion of the vector field is given by
\begin{equation}
\frac{d}{dt} \left[f^2 e^{\alpha +4\sigma}\dot{v}\right]=0,
\end{equation}
which can be solved as
\begin{equation}
 \dot{v}=\frac{e^{-\alpha-4\sigma}}{f(\phi)^2}c_{A},    \label{v}
\end{equation}
where $c_A$ is the constant of integration and 'dot' represents derivative with respect to time.\\
Substituting (\ref{metric-aniso}), (\ref{v}) in (\ref{grav_eqn}) and (\ref{box_scalar}) 
together with (\ref{GBform}) and  $F_{\mu \nu}$,
we obtain gravitational field equations and equation of motion of the inflaton field as follows 
 \begin{eqnarray}
 \dot{\alpha}^2 &=& \dot{\sigma}^2 +\frac{\kappa^2}{3} \left[V(\phi) +\frac{\dot{\phi}^2}{2}+
 \frac{e^{-4\alpha-4\sigma}}{2 f(\phi)^2}c_{A}^2\right]
 -\kappa^2 \dot{\xi}(\dot{\alpha}-2 \dot{\sigma})(\dot{\alpha}+ \dot{\sigma})^2  \label{ttcomp} ,\\
\ddot{\sigma}  &=& -3\, \dot{\alpha}\, \dot{\sigma}+
 \kappa^2 \displaystyle\frac{c_{A}^2}{3 f(\phi)^2}e^{-4\alpha-4\sigma} \label{aniso-comp} \\  [3mm]  \nonumber
&&  - \kappa^2 \dot{\xi}\displaystyle\left[\dot{\alpha}(3 \dot{\sigma}^2 + \ddot{\sigma})+\dot{\sigma}\left(\ddot{\alpha}+
 2 \ddot{\sigma}\right)+
 3\dot{\alpha}^2\dot{\sigma}\right]- \kappa^2\ddot{\xi} (\dot{\alpha}\dot{\sigma}+\dot{\sigma}^2), \\  
\ddot{\alpha} &=& -3\dot{\alpha}^2 +\kappa^2 \left[ V(\phi) + 
 \displaystyle\frac{c_{A}^2}{6 f(\phi)^2}e^{-4\alpha-4\sigma} \right] \label{scale-comp}\\[3mm]   \nonumber
 && + \displaystyle\frac{\kappa^2}{2} \dot{\xi} \left[-5\dot{\alpha}^3+ \dot{\alpha}(9\dot{\sigma}^2-2\ddot{\alpha})+
4 \dot{\sigma}^3+2\dot{\sigma}\ddot{\sigma}\right] 
+\displaystyle \frac{\kappa^2}{2}\ddot{\xi}(-\dot{\alpha}^2 + \dot{\sigma}^2),  \\
\ddot{\phi} & = & -3\,\dot{\alpha}\,\dot{\phi}-V'(\phi) +
 \displaystyle \frac{c_{A}^2 f'(\phi) e^{-4\alpha -4\sigma}}{f(\phi)^3}   \label{scalar_comp} \\[3mm]  \nonumber
 && + 3\xi' (\dot{\alpha}+ \dot{\sigma})\left[\dot{\alpha}^3-\dot{\alpha}^2\dot{\sigma}+
 \dot{\alpha}(-2 \dot{\sigma}^2+\ddot{\alpha})-\dot{\sigma}(\ddot{\alpha}+ 2\ddot{\sigma})\right], 
\end{eqnarray}
where $\dot{\xi}=\xi'(\phi)\dot{\phi}$ and $\ddot{\xi}=\xi''(\phi) \dot{\phi}^2+\xi'(\phi)\ddot{\phi}$.
\\[1mm]
Let us now define the Hubble's expansion rate as $H \equiv \dot{\alpha}$, 
then (\ref{ttcomp}) can be expressed as
\begin{equation}
 H^2 = \dot{\sigma}^2 
 +\frac{\kappa^2}{3} \left[V(\phi) +\frac{\dot{\phi}^2}{2}+\frac{e^{-4\alpha-4\sigma}}{2 f(\phi)^2}c_{A}^2\right]
 -\kappa^2 \dot{\xi}(H-2 \dot{\sigma})(H+ \dot{\sigma})^2,    \label{Ham}
\end{equation}
Here, contributions of the vector field and the Gauss-Bonnet term add as effective
potential in addition to the inflaton potential $V(\phi)$ and influence the dynamics of the inflation.
The energy density of the vector field can be defined as
\begin{equation}
 \rho_{v}=\displaystyle\frac{e^{-4\alpha-4\sigma}}{2 f(\phi)^2}c_{A}^2  \label{rho},
\end{equation}
The conventional slow-roll inflation continues when $\dot{\sigma}$ is suppressed over H i,e.
$\displaystyle\frac{\dot{\sigma}}{H}< <1$ 
and two slow-roll conditions $\, \displaystyle\frac{1}{2}\dot{\phi}^2 << V(\phi)$, $\ddot{\phi}<<3H \dot{\phi}$
and additionally $V(\phi)>>\rho_v$ are satisfied.
In presence of scalar coupled Gauss-Bonnet term, 
we have two additional slow-roll conditions i,e. $\dot{\xi}H<<1$ and 
$\ddot{\xi}<<\dot{\xi}H$ which hold during inflation.
Under these conditions, (\ref{Ham}) yields the Friedmann equation 
\begin{equation}
 H^2\simeq \kappa^2\displaystyle \frac{V(\phi)}{3},   \label{Friedmann}
\end{equation}
which leads to the accelerated expansion of the universe when the inflaton potential remains nearly constant.
The equation of motion of the scalar field under slow-roll conditions becomes
\begin{equation}
 3H \dot{\phi} + V'(\phi) - 3\xi'H^4\simeq 0.
\end{equation}
From these two relations one obtains,
\begin{equation}
 \displaystyle \frac{d\alpha}{d \phi} = \displaystyle \kappa^2 \frac{3V}{(-3V'+\kappa^4 \xi'V^2)} = 
 \displaystyle \frac{\kappa^2}{Q},    \label{Q1}
\end{equation}
where $Q=\displaystyle\frac{-3V'+\kappa^4\xi'V^2}{3V}$.\\
Thus for sustaining slow-roll inflation $V(\phi)>> \rho_{v}$. However the energy density of the 
vector field $\rho_{v}$ may still remain constant and may contribute to anisotropic effects. 
Under the approximation $\sigma <<\alpha$, $\rho_{v}$ can remain almost constant
if the functional form of $f(\phi)$ from (\ref{rho}) can be expressed in the following form 
\begin{equation}
 f(\phi)=e^{-2\alpha}.
\end{equation}
More generally,  we may parametrize $f(\phi)$ such that
\begin{equation}
  f(\phi)=e^{-2 c \alpha},
\end{equation}
where, $c$ is a constant parameter. 
Then, in the slow-roll phase, the functional form of $f(\phi)$ using (\ref{Q1}) becomes
\begin{equation}
 f(\phi) = e^{-2c \kappa^2 \int \frac{3 V}{-3V'+\kappa^4 \xi'V^2}d \phi}.   \label{f}
\end{equation}
This shows that the general form of the gauge coupling function is related to both the inflaton potential and 
the Gauss-Bonnet coupling function and thus $\rho_{v}$ also depends on quadratic curvature corrections through 
Gauss-Bonnet term. 
We note that the functional form of $f(\phi)$ exactly reduces
to the non- Gauss-Bonnet result when $\xi(\phi)=\xi'(\phi)=0$ \cite{Watanabe}.
So long as $\rho_{v}$ is negligibly small, conventional isotropic Gauss-Bonnet inflation
is realized. 
However, as $\rho_{v} \propto e^{4(c-1)\alpha}$, the vector field grows during inflation when $c>1$ supported by Gauss-Bonnet corrections. 
 Although $\rho_{v}$  can be neglected in (\ref{Friedmann}) compared to inflaton potential, it can contribute non-trivially in 
the scalar field equation of motion to genetate anisotropy in the $c>1$ regime.  
Using (\ref{f}) we arrive at a condition
\begin{equation}
 \displaystyle \frac{1}{6\kappa^2}\displaystyle \frac{f'(3 V'-\kappa^4 \xi'V^2)}{f V} >1,  \label{cond1}
\end{equation}
which suggests that a vector hair can exist during the slow-roll phase of inflation in the Gauss-Bonnet gravity scenario provided any given set of
$f(\phi)$, $V(\phi)$ and $\xi(\phi)$ satisfies the condition (\ref{cond1}).
Assuming that anisotropy attains a constant value such that $\ddot{\sigma}\approx 0$ in the anisotropic phase, the inflaton dynamics can be studied from the the equation of motion of the inflaton field (\ref{scalar_comp}) which is given by
\begin{equation}
 \ddot{\phi}= -3H\dot{\phi}-V'+\displaystyle\frac{c_A^2 f'(\phi)}{f^3(\phi)}e^{-4\alpha-4\sigma}+ 
 3\xi' H^4 \left(1+\frac{\dot{H}}{H^2}-
 3 \frac{\Sigma^2}{H^2}-2 \frac{\Sigma^3}{H^3}- \frac{\Sigma^2 \dot{H}}{H^4}\right),
\end{equation}
where we have defined $\Sigma\equiv \dot{\sigma}$. 
Since in general anisotropic effects are small, we neglect higher orders of $\displaystyle\frac{\Sigma}{H}$ and 
the above equation using (\ref{f}) reduces to
\begin{equation}
  \ddot{\phi}= -3H\dot{\phi}+ 3\xi' H^4 \left(1+\frac{\dot{H}}{H^2}\right)
  -V'\left(1-\frac{2c}{\epsilon_v}{\cal {R}}\right),   \label{1}
\end{equation}
where we define ${\cal {R}} = \displaystyle\frac{\rho_{v}}{V(\phi)}$ and the
slow-roll parameter is $\epsilon_{v}=-\displaystyle \frac{1}{2\kappa^2}\frac{V'}{V}Q$. \\
 With  the expansion of the universe, $\rho_{v}$ also increases with $\alpha$ and can no longer be neglected.
 From the scalar field equation, we observe that as universe expands,  anisotropy is appreciable when ${\cal{R}}=\displaystyle\frac{\rho_{v}}{V(\phi)}$ 
is comparable to $\displaystyle\frac{\epsilon_v}{2c}$. 
If now during anisotropic phase,  $\rho_{v}$ increases further such that ${\cal{R}}>>\displaystyle\frac{\epsilon_v}{2c}$ i,e. $\displaystyle \frac{2c}{\epsilon_{v}} {\cal R}>>1$, the
inflaton field never rolls down as can be seen from (\ref{1})
and hence $\rho_v$ decreases so that inflation resumes again \cite{Watanabe}. Therefore $ \rho_{v}<< V(\phi)$ always remains valid such that inflation occurs while the vector field keeps contributing to anisotropy.\\
As  $\rho_v$ grows in the slow-roll inflation regime characterized by approximations 
$\ddot{\phi}<<3H\dot{\phi}$ and $\dot{H}<<H^2$,
the inflaton dynamics can be captured from the scalar field equation
\begin{equation}
3 H \dot{\phi}= -V' + \displaystyle\frac{c_A^2 f'(\phi)}{f^3(\phi)}e^{-4\alpha-4\sigma} \label{sc-dyn}                                                           +3\xi'H^4 ,
\end{equation}
where we have assumed higher orders of $\displaystyle \frac{\Sigma}{H}$ are negligibly small.
Now dividing (\ref{sc-dyn}) by $3H^2$ and using (\ref{rho}), (\ref{Friedmann}) and (\ref{f}) 
in presence of Gauss-Bonnet gravity we obtain
\begin{equation}
 \displaystyle \frac{d\phi}{d\alpha} = \displaystyle\frac{-3V'+\kappa^4\xi'V^2}{3 \kappa^2 V} +
 \displaystyle \frac{6 c_A^2\,c}{3 V'-\kappa^4\xi'V^2}e^{-4\alpha -4\sigma-4\kappa^2 c\int\frac{3 V}{3V'-\kappa^4\xi'V^2} d \phi }.
\end{equation}
We now integrate the above equation and neglect the variations of $V$,$V'$ and $\sigma$ with respect to $\alpha$, so that
\begin{equation}
e^{4\alpha +4\sigma+4\kappa^2 c\int\frac{3 V}{3V'-\kappa^4\xi'V^2} d \phi} =
\displaystyle \frac{6 c^2 c_{A}^2}{c-1}\displaystyle \frac{3 \kappa^2 V}{(3V'-\kappa^4\xi'V^2)}
(1+A e^{-4(c-1)\alpha-4\sigma}),   \label{soln} 
\end{equation}
where $A$ is the constant of integration. 
Using (\ref{f}), its derivative and (\ref{soln}) in the scalar field equation, we get, 
\begin{equation}
 \displaystyle \frac{d\phi}{d \alpha}=\displaystyle\frac{-3V'+\kappa^4\xi'V^2}{3 \kappa^2 V}+
 \displaystyle \frac{c-1}{c}\left[\frac{3V'-\kappa^4\xi'V^2}{3 \kappa^2 V}(1+A e^{-4(c-1)\alpha-4\sigma})^{-1}\right]. 
\end{equation}
In the limit $\alpha \rightarrow -\infty$, ${(1+A e^{-4(c-1)\alpha-4\sigma})}^{-1} \rightarrow 0$, so that  conventional slow-roll inflation is encountered as described by,
\begin{equation}
 \displaystyle \frac{d\phi}{d \alpha}= \displaystyle\frac{-3V'+\kappa^4\xi'V^2}{3 \kappa^2 V}=\displaystyle \frac{Q}{\kappa^2}.  \label{Q}
\end{equation}
At this point, the energy density of the vector field $\rho_{v} \rightarrow 0$ thus suggesting vanishingly small contribution to the slow-roll inflation. On the other hand,
as  $\alpha \rightarrow  \infty$, the quantity ${(1+A e^{-4(c-1)\alpha-4\sigma})}^{-1} \rightarrow 1$. 
Therefore we can write,
\begin{equation}
 \displaystyle \frac{d\phi}{d \alpha}=\displaystyle \frac{1}{c} \displaystyle\frac{-3V'+\kappa^4\xi'V^2}{3 \kappa^2 V} = \displaystyle \frac{1}{c}\frac{Q}{\kappa^2}.
 \label{A}
\end{equation}
This relation as compared to (\ref{Q}) is now modified during the 
inflationary phase which is now accompanied with small but non-zero effects of anisotropy. 
We observe that in the modified slow-roll phase (\ref{A}) is $1/c$ times reduced compared to (\ref{Q}). 
Now the energy density of the vector field can be expressed as,
\begin{equation}
 \rho_{v}=\displaystyle \frac{c-1}{c^2}\epsilon_{v}\frac{V(3V'-\kappa^4\xi'V^2)}{6V'}.  \label{vec_en}
\end{equation}
This relation suggests that the Gauss-Bonnet term contributes non-trivially to the energy density of the vector field where $c>1$ must be taken.
Let us now consider the anisotropy equation given by (\ref{aniso-comp}).
Since during the anisotropic phase, the anisotropy is assumed to attain a constant value, 
we assume $\ddot{\sigma}<<\dot{\alpha}\dot{\sigma}$ and $\dot{\Sigma}\approx 0$.
Then under slow-roll conditions $\dot{\xi}H<<1$ and $\ddot{\xi}<<\dot{\xi}H$,
the anisotropy equation may be written as, 
\begin{equation}
3 H \Sigma + \kappa^2 \dot{\xi} (3 H \Sigma^2+\Sigma \dot{H}+3H^2\Sigma)=\displaystyle\frac{2\kappa^2}{3}\rho_{v}.
\end{equation}
Dividing the above equation by $3H^2$, we have,
\begin{equation}
 \displaystyle \frac{\Sigma}{H}+\kappa^2 \dot{\xi}H\left[\displaystyle\frac{\Sigma^2}{H^2}+
 \displaystyle\frac{\Sigma \dot{H}}{3H^3}+\displaystyle \frac{\Sigma}{H}\right]=\displaystyle \frac{2\rho_{v}}{3V(\phi)}.
 \end{equation}
Now $\dot{H}<<H^2$ and since $\Sigma$ is suppressed over $H$, we assume that 
$\displaystyle\frac{\Sigma^2}{H^2}<<\displaystyle\frac{\Sigma}{H}$, so that we have
\begin{equation}
 \displaystyle\frac{\Sigma}{H}(1+\delta_{H})=\displaystyle \frac{2\rho_{v}}{3V(\phi)} =\frac{2}{3}{\cal R},
\end{equation}
where the extra slow-roll parameter  $\delta_{H}=\kappa^2 \dot{\xi}H$ appears only in presence of Gauss-Bonnet correction.
Now substituting (\ref{vec_en}) in the above equation, we obtain
\begin{equation}
 \displaystyle\frac{\Sigma}{H}(1+\delta_{H})=\displaystyle\frac{1}{3}\frac{c-1}{c^2}
 (\epsilon_{v}+\displaystyle \frac{1}{2}\delta_{v}),
\end{equation}
where we define $\delta_{v}=\displaystyle\frac{\kappa^2}{3} \xi' V Q$.\\
Let us now substitute (\ref{ttcomp}) in the scale-factor equation given by (\ref{scale-comp}).
Under the approximations $\dot{\sigma}^2 << \frac{1}{2}\dot{\phi}^2$, $\dot{\Sigma}\approx0$, $\ddot{\xi}<<\dot{\xi}H$
and neglecting higher powers of $\displaystyle\frac{\Sigma}{H}$ greater than $1$, the scale factor equation becomes,
\begin{equation}
 \ddot{\alpha}= -\frac{1}{2}\kappa^2\dot{\phi}^2-\frac{1}{3}\kappa^2 c_A^2 \frac{e^{-4\alpha-4\sigma}}{f(\phi)^2}
 +\frac{\kappa^2}{2}\dot{\xi}\dot{\alpha}^3.
\end{equation}
Since $- \displaystyle \frac{\ddot{\alpha}}{\dot{\alpha}^2}=\epsilon_{H}$, therefore the above equation 
can be written as,
\begin{equation}
\epsilon_{H}= \displaystyle\frac{1}{2}\kappa^2\displaystyle \left(\frac{d\phi}{d\alpha}\right)^2 +
 c_A^2 \frac{e^{-4\alpha-4\sigma}}{f(\phi)^2 V}
 -\frac{1}{2}\delta_{H}.
\end{equation}
Using (\ref{A}) and (\ref{rho}), the above expression using (\ref{Friedmann}) becomes,
\begin{equation}
 \epsilon_{H}+\frac{1}{2}\delta_{H}=\displaystyle \frac{1}{c}(\epsilon_{v}+\displaystyle \frac{1}{2}\delta_{v}).
\end{equation}
Then the measure of anisotropy is given by,
\begin{equation}
\displaystyle \frac{\Sigma}{H}(1+\delta_{H})=\displaystyle \frac{1}{3}\frac{c-1}{c}(\epsilon_{H}+\frac{1}{2}\delta_{H}). 
\end{equation}
As $\epsilon_{H}$ and $\delta_{H}$ must be very small to sustain inflation, 
therefore up to first order, we have the following relation, 
\begin{equation}
\displaystyle \frac{\Sigma}{H}=\displaystyle \frac{1}{3}\frac{c-1}{c}(\epsilon_{H}+\frac{1}{2}\delta_{H}). \label{expr_c}  
\end{equation}
Therefore when Gauss-Bonnet corrections are taken into consideration where the Gauss-Bonnet coupling 
is a function of inflaton field, we find that the anisotropy is proportional to slow-roll parameters of 
the theory namely $\epsilon_H$ and $\delta_{H}$. 
We note that in absence of Gauss-Bonnet correction (\ref{expr_c}) reduces to the result obtained in \cite{Watanabe}.

\section{Power law inflation with Gauss-Bonnet gravity : Exact solutions }
After presenting a general set-up for realizing anisotropic inflation with Gauss-Bonnet correction term,
we now aim to construct exact anisotropic power-law solutions.
These solutions are obtained by taking combined effects of the inflaton potential, the gauge kinetic function and
the Gauss-Bonnet coupling function.
We assume that inflaton potential is exponentially dependent on 
the inflaton field such that,
\begin{equation}
 V(\phi)=V_{0} e^{\kappa \lambda \phi}.    \label{infl_potential}
\end{equation}
 Similarly, we also assume that gauge coupling and Gauss-Bonnet coupling are exponential functions
 of the inflaton field  
\begin{equation}
f(\phi)=f_0 e^{ \kappa \rho \phi} \qquad  \xi(\phi)= \xi_{0}e^{\kappa \omega \phi},  \label{infl_coupling}
\end{equation}
where $V_{0}, f_{0}, \xi_{0},\lambda,\rho$ and $\omega$ are constant parameters of the theory.
\subsection{Isotropic inflation}
Let us first discuss the isotropic power-law solution in presence of Gauss-Bonnet term
for which we assume the following metric
\begin{equation}
 ds^2=-dt^2 + e^{2\alpha(t)}(dx^2+dy^2+dz^2),  \label{iso_metric}
\end{equation}
where $e^{\alpha(t)}$ is the isotropic scale factor. 
The isotropic inflation corresponds to no anisotropy and hence 
inconsequential contributions of the vector field.
Substituting the above metric in (\ref{grav_eqn})-(\ref{box_scalar}), the gravitational field equations
and the equation of motion of the inflaton field are as follows
\begin{eqnarray}
 \dot{\alpha}^2  & = &\displaystyle\frac{\kappa^2}{6}\dot{\phi}^2 +
 \displaystyle\frac{\kappa^2}{3}V(\phi)- \kappa^2\dot{\xi}\dot{\alpha}^3,    \label{iso1}\\
\ddot{\alpha} & =& -3\dot{\alpha}^2+ \kappa^2 V(\phi)-
\frac{\kappa^2}{2}\dot{\xi}\left(5\dot{\alpha}^3+2 \dot{\alpha}\ddot{\alpha}\right)-
\displaystyle\frac{\kappa^2}{2}\ddot{\xi}\dot{\alpha}^2,    \label{iso2}\\
\ddot{\phi}& =&-3\dot{\alpha}\dot{\phi}-V'(\phi)+3\dot{\alpha}^2\xi' (\dot{\alpha}^2+\ddot{\alpha}),  \label{iso3}
\end{eqnarray}
where $\dot{\xi}=\xi'(\phi)\dot{\phi}$ and $\ddot{\xi}=\xi''(\phi) \dot{\phi}^2+\xi'(\phi)\ddot{\phi}$.\\
To find power-law isotropic solutions, we consider following ansatz
\begin{equation}
 \alpha=\zeta \ln t,\qquad \kappa \phi=\chi \ln t+\phi_0,     \label{iso_scale}
 \end{equation}
where $\phi_{0}$ is a constant quantity.
The inflaton potential and Gauss-Bonnet coupling functions are 
\begin{equation}
 V(\phi)=V_0 e^{\lambda \kappa \phi},\quad \xi(\phi)=\xi_{0} e^{\omega \kappa \phi}.   \label{iso_pot}
\end{equation}
Substituting (\ref{iso_scale}) and (\ref{iso_pot}) in (\ref{iso1})- (\ref{iso3}) following
conditions are obtained 
\begin{eqnarray}
 \lambda \chi=-2,\quad \chi \omega =2  \label{iso_cons0},
 \end{eqnarray}
which imply $\lambda=-\omega$.
From (\ref{iso1}) balance of amplitudes gives
\begin{equation}
 -\zeta^2+\frac{1}{3}u+\frac{1}{6}\chi^2-2n \zeta^3=0,  \label{iso_cons1}
\end{equation}
where we have defined $u=\kappa^2 V_0 e^{\kappa \lambda \phi_0}$ and $n=\kappa^2 \xi_0 e^{\kappa \omega \phi_0}$.\\
We denote $n$ as the Gauss-Bonnet parameter such that $n\rightarrow 0$ limit leads to the non-Gauss-Bonnet case.
The balance of amplitudes in (\ref{iso2}) gives,
\begin{equation}
 \zeta -3\zeta^2 + u - n (5\zeta^3-2\zeta^2)-n\zeta^2=0.   \label{iso_cons2}
\end{equation}
Then substituting $u$ from (\ref{iso_cons1}) in (\ref{iso_cons2}) together with (\ref{iso_cons0}), we obtain an equation for $\zeta$ as,
\begin{equation}
 \zeta+n(\zeta^2+\zeta^3)-\frac{2}{\lambda^2}=0.
\end{equation}
This is a cubic polynomial equation of $\zeta$ which on solving gives three roots out 
of which two are imaginary and one is real. The real root of $\zeta$ is given by,  
\begin{eqnarray}
\zeta &= & -\frac{1}{3}+\frac{2^{1/3} \lambda ^2 (n-3)}{3 \left(3 \sqrt{3} \sqrt{\lambda ^8 n^3 
\left(\lambda ^4 (4-n)+4 \lambda ^2 (9-2 n)
   n+108 n\right)} +\lambda ^4 n^2 \left(\lambda ^2 (9-2 n)+54\right)\right)^{1/3}} \nonumber \\[2mm]
   && + \frac{2^{2/3} \left(3 \sqrt{3} \sqrt{\lambda ^8 n^3 
\left(\lambda ^4 (4-n)+4 \lambda ^2 (9-2 n)
   n+108 n\right)} +\lambda ^4 n^2 \left(\lambda ^2 (9-2 n)+54\right)\right)^{1/3}}{ 6 \lambda ^2 n}  \label{iso_zeta}.
   \nonumber\\
\end{eqnarray}
In the leading order of $n$ plus higher order terms in $n$, the solution of $\zeta$ can be expressed as,
\begin{equation}
 \zeta = \frac{2}{\lambda^2}-\frac{4}{\lambda^6}(2+\lambda^2)n + \mathcal{O}(n^2)+\mathcal{O}(n^3) +..   \label{fin_iso_zeta}
\end{equation}
Then the metric solution for isotropic inflation becomes
\begin{equation}
ds^2=-dt^2 + t^{2\zeta}(dx^2 + dy^2 + dz^2), 
\end{equation}
where $\zeta$ is given by (\ref{fin_iso_zeta}). However, for sufficiently fast power-law inflation we
demand $\zeta>>1$. 
Then in the leading order of $n$, the following condition
\begin{equation}
 \frac{4}{\lambda^6}(2+\lambda^2)n<<\frac{2}{\lambda^2}   \label{cond}
\end{equation}
must be true.
This condition suggests that $n<<1$  and can be treated as a correction term if $\lambda<<1$. 
As an estimation, if $\lambda=0.1$ is taken then (\ref{cond}) implies $n<<10^{-4}$ on the other hand $\lambda>>1$
implies $n>>1$, the situation which is not desirable. 
So it is sufficient to consider terms in $\zeta$ up to leading order in $n$.  
We further note from (\ref{fin_iso_zeta}) that in the limit $n\rightarrow 0$,
$\zeta=\displaystyle\frac{2}{\lambda^2}$ is exactly reproduced as obtained in \cite{Kanno} in absence of Gauss-Bonnet
corrections.
\subsection{Anisotropic inflation}
We now look forward to the construction of anisotropic power-law solutions in presence of Gauss-Bonnet correction with 
the help of a non-trivial vector field. 
With our choice of gauge the vector field can be expressed as $A_{\mu} dx^{\mu}=v(t)d x$. 
The potential $V(\phi)$ of the inflaton field $\phi$, the gauge coupling function and 
the Gauss-Bonnet coupling assume the form given by (\ref{infl_potential}) and (\ref{infl_coupling}) respectively.
Since $x-$ axis is taken as the direction of the vector field, there exists a rotational symmetry in the 
$y-z$ plane. 
For studying anisotropic power law solution, we consider the following metric
\begin{equation}
 ds^2 = -dt^2 +e^{2\alpha(t)}\left[ e^{-4\sigma(t)}dx^2 + e^{2\sigma(t)}(dy^2 + dz^2) \right],  \label{metric_aniso}
\end{equation}
and assume following ansatz for power-law anisotropic solutions and the inflaton field
\begin{eqnarray}
\alpha(t) = \zeta \ln t    \label{alpha},\\[1mm]
\sigma(t) =\eta \ln t    \label{sigma},  \\[1mm]
\kappa \phi(t) = \chi \ln t + \phi_{0},   \label{phi}  
\end{eqnarray}
where $\phi_{0}$ is a constant quantity. 
Substituting these in (\ref{ttcomp}) and comparing the powers of $t$,  we obtain following conditions
\begin{eqnarray}
 \rho \chi+ 2 \zeta + 2\eta &=& 1,    \label{cons_1} \\[1mm]
 \lambda \chi &=& -2,                  \label{cons_2} \\[1mm]
 \chi \omega &=& 2.                           \label{cons_3}
\end{eqnarray}
The last two conditions suggest $\lambda = -\omega$.
The balance of amplitudes in (\ref{ttcomp}) gives
\begin{equation}
 -\zeta^2 + \eta^2 + \displaystyle\frac{1}{6} \chi^2 +\displaystyle \frac{1}{3}u +\frac{1}{6} v
 -n \chi \omega (\zeta - 2 \eta)(\zeta + \eta)^2 = 0,   \label{Hamilton}
\end{equation}
where we define $u = \kappa^2 V_{0} e^{\lambda \phi_{0}}, \,v =\kappa^2 \displaystyle \frac{e^{-2\rho \phi_{0}}}{f_{0}^2} c_{A}^2$ and
$n = \kappa^2\xi_{0}\, e^{\omega \phi_{0}}$. 
Here $n$ signifies the contribution of Gauss-Bonnet term.\\
 The anisotropy equation given by (\ref{aniso-comp}) simultaneously satisfying (\ref{cons_1})-(\ref{cons_3}) yields,
\begin{equation}
 -\eta + 3 \zeta \eta -\frac{1}{3} v +n \chi \omega \eta (\zeta + \eta)(-3 +3 \zeta +\chi \omega) =0.   \label{aniso}
\end{equation}
The scale factor equation (\ref{scale-comp}) reproducing (\ref{cons_1})-(\ref{cons_3}) leads to
\begin{equation}
 \zeta -3 \zeta^2 + u +\frac{1}{6}v - \frac{n}{2}\chi \omega (\zeta + \eta ) \left[5\zeta^2 + \zeta (-3-5 \eta +\chi \omega)
 -\eta(-3+4\eta+\chi \omega) \right] = 0.   \label{scale}
\end{equation}
Similarly the equation of motion of the scalar field $\phi$ gives,
\begin{equation}
 -\chi +3 \zeta \chi +\lambda u-\rho v -3 n \omega (-1+\zeta)(\zeta -2 \eta)(\zeta +\eta)^2 = 0.   \label{scalar}
\end{equation}
From (\ref{aniso}), (\ref{scale}) and using the constraint equations, we can express $u$ and $v$ as
\begin{eqnarray}
u &=&\frac{1}{4}+\frac{9 \zeta ^2}{2}-\frac{3 \zeta  \rho }{2 \lambda }-\frac{9 \zeta }{4}+\frac{\rho }{2 \lambda }+
\frac{n \left(9 \zeta ^2 \lambda ^3+
18 \zeta^2 \lambda ^2 \rho -3 \zeta  \lambda ^3-6 \zeta  \lambda ^2 \rho -2 \lambda ^2 \rho -
8 \lambda  \rho ^2-8 \rho ^3\right)}{2 \lambda^3}, \nonumber \label{new_u} \\[2mm]
v &=& -\frac{3}{2} -9 \zeta ^2+\frac{9 \zeta  \rho }{\lambda }+\frac{15 \zeta }{2}-\frac{3 \rho }{\lambda }-
\frac{3 n (3 \zeta -1) (\lambda +2 \rho ) \left[(2 \zeta -1)\lambda -2 \rho \right]}{2 \lambda ^2}
 \end{eqnarray}
Using (\ref{cons_1})-(\ref{cons_3}) and substituting $u$, $v$ in the scalar field equation and  we get
\begin{equation}
\frac{(3 \zeta -1) \left[(\lambda +2 \rho ) (\lambda(6 \zeta -1) -6 \rho )-8\right]}{4 \lambda }+
\frac{n(3 \zeta -1) (\lambda +2 \rho )^2 ((9 \zeta -3) \lambda -8 \rho )}{4 \lambda ^2}=0,
\end{equation}
which can be solved for $\zeta$ to obtain
\begin{equation}
 \zeta = \displaystyle\frac{1}{3},\quad \zeta= \frac{\lambda ^3 (3 n+1)+4 \lambda ^2 (5 n+2) \rho +4 \lambda  \left((11 n+3) \rho ^2+2\right)+32 n \rho ^3}{3 \lambda  (\lambda +2 \rho )
   (\lambda  (3 n+2)+6 n \rho )}.
\end{equation}
Out of these two solutions, we discard $\zeta= \displaystyle\frac{1}{3}$ because substituting it back in 
$v$ gives $v=0$ which implies $c_{A}=0$. 
Hence the second solution of $\zeta$ is taken which is rearranged as
\begin{equation}
\zeta=\frac{\lambda ^3+8 \lambda ^2 \rho +12 \lambda  \rho ^2+8 \lambda +
n \left(3 \lambda ^3+20 \lambda ^2 \rho +44 \lambda  \rho ^2+32 \rho^3\right)}{6 \lambda ^3+12 \lambda ^2 \rho +
n \left(9 \lambda ^3+36 \lambda ^2 \rho +36 \lambda  \rho ^2\right)}  \label{zeta}
\end{equation}
and using (\ref{cons_1}) and (\ref{cons_2}) we obtain 
\begin{equation}
\eta=\frac{4 \lambda ^3+8 \lambda ^2 \rho -16 \lambda +n \left(3 \lambda ^3+
14 \lambda ^2 \rho +20 \lambda  \rho ^2+8 \rho ^3\right)}{12 \lambda^3+24 \lambda ^2 \rho +
n \left(18 \lambda ^3+72 \lambda ^2 \rho +72 \lambda  \rho ^2\right)}.  \label{eta}
\end{equation}
It is to be noted that in absence of Gauss-Bonnet term, $\zeta$ and $\eta$ reduce to their respective same forms 
as obtained in \cite{Kanno}.
So anisotropic power-law solutions exist in the context of Gauss-Bonnet gravity and 
the corresponding metric becomes
\begin{equation}
 ds^2=-dt^2 + t^{2\zeta -4 \eta}dx^2 +t^{2\zeta +2 \eta}(dy^2 +dz^2), 
\end{equation}
where $\zeta$ and $\eta$ are given by (\ref{zeta}) and (\ref{eta}) respectively.\\
In terms of $\rho$, $\lambda$ and $n$ we can express $u$ and $v$ as follows 
\begin{equation}
u=\displaystyle \frac{a_0+n a_1+ n^2 a_2+n^3 a_3 }{b_0+n b_1 + n^2 b_2 },  
\end{equation}
where $a_0=-8 \lambda ^3 ((\lambda -6 \rho ) (\lambda +2 \rho )-8) (\rho  (\lambda +2 \rho )+2)$,\\
$a_1=-\lambda ^2 (\lambda +2 \rho ) \left[(\lambda +2 \rho ) \left(5 \lambda ^3+18 \lambda ^2 \rho -4 \lambda  \left(37 \rho ^2+6\right)-8 \rho 
   \left(41 \rho ^2+50\right)\right)-128\right]$,\\
$a_2=-2 \lambda  (\lambda +2 \rho )^3 \left[3 \lambda ^3-
4 \lambda  \left(25 \rho ^2+6\right)-16 \rho  \left(11 \rho ^2+8\right)\right]$,
$a_3=4 \rho  (\lambda +2 \rho )^5 (3 \lambda +14 \rho )$,\\
$b_0=16 \lambda ^5 (\lambda +2 \rho )^2$,
$b_1=48 \lambda ^4 (\lambda +2 \rho )^3$, $b_2=36 \lambda ^3 (\lambda +2 \rho )^4$\\ 
and
\begin{equation}
 v=\displaystyle \frac{c_0+n c_1+ n^2 c_2+n^3 c_3 }{d_0+n d_1 + n^2 d_2 }, \label{v1}
\end{equation}
where $c_0=-4 \lambda ^3 \left(\lambda ^2+2 \lambda  \rho -4\right) [(\lambda -6 \rho ) (\lambda +2 \rho )-8]$,\\
$c_1=-\lambda ^2 (\lambda +2 \rho ) \left[(\lambda +2 \rho ) \left(7 \lambda ^3-58 \lambda ^2 \rho -
12 \lambda  \left(13 \rho ^2+6\right)+8 \rho 
   \left(26-3 \rho ^2\right)\right)+128\right]$,\\
$c_2=-\lambda  (\lambda +2 \rho )^3 \left[3 \lambda ^3-66 \lambda ^2 \rho -4 \lambda  \left(43 \rho ^2+6\right)-56 \rho  \left(\rho
   ^2-2\right)\right]$,
$c_3=8 \rho  (\lambda +2 \rho )^5 (3 \lambda +2 \rho )$.\\
$d_0=8 \lambda ^5 (\lambda +2 \rho )^2$, $d_1=24 \lambda ^4 (\lambda +2 \rho )^3$, $d_2=18 \lambda ^3 (\lambda +2 \rho )^4$.\\[1mm]
It is to be noted that in the limit $n\rightarrow 0$, the expressions of $u$ and $v$ 
match with results obtained for non-Gauss Bonnet case \cite{Kanno}.
In the given scenario, the anisotropy can be measured from the quantity
\begin{equation}
\displaystyle \frac{\Sigma}{H}=\displaystyle \frac{\eta}{\zeta}=\frac{4 \lambda ^3+8 \lambda ^2 \rho -16 \lambda +
 n \left(3 \lambda ^3+14 \lambda ^2 \rho +20 \lambda  \rho ^2+8 \rho ^3\right)}{2 \lambda ^3+16
\lambda ^2 \rho +24 \lambda  \rho ^2+16 \lambda +n \left(6 \lambda ^3+40 \lambda ^2 \rho +
88 \lambda  \rho ^2+64 \rho ^3\right)}.   \label{expr_aniso}
\end{equation}
Now (\ref{zeta}) and (\ref{eta}) suggest that for power-law anisotropic inflation to occur, $\rho>>1$ while $\lambda <<1$
and $n<<1$.
We emphasize here that the positivity of $n$ is a requirement in our analysis.
As $v=\displaystyle \frac{\kappa^2 e^{-2 \rho \phi_{0}}}{f_0 ^2}c_{A}^2$ is defined as a squared quantity, 
(\ref{v1}) shows that even a small negative value of Gauss-Bonnet parameter $n$ leads to $v<0$
over a large range of $\lambda$ and $\rho$.
Since $v$ cannot be negative for any value of $\lambda$ and $\rho$, therefore we will always consider $n>0$.\\
The quantity $\displaystyle\frac{\Sigma}{H}$ exactly reduces to the 
same relation obtained in \cite{Kanno} when $n=0$ is substituted in (\ref{expr_aniso}). 
Therefore, our analysis gives a scope to compare anisotropy i,e. $\Sigma/H$ between 
$n=0$ case and the situation when the Gauss-Bonnet term is turned on.
The following plot for anisotropy vs $n$ clearly shows
that for $n>0$, Gauss-Bonnet correction enhances the anisotropy during inflation.
From figure 1, we find that as $\rho$ is increased, 
anisotropy also increases  which is showed for three different values of $\rho$ and for different values of $n$.
\begin{figure}[ht]
\includegraphics{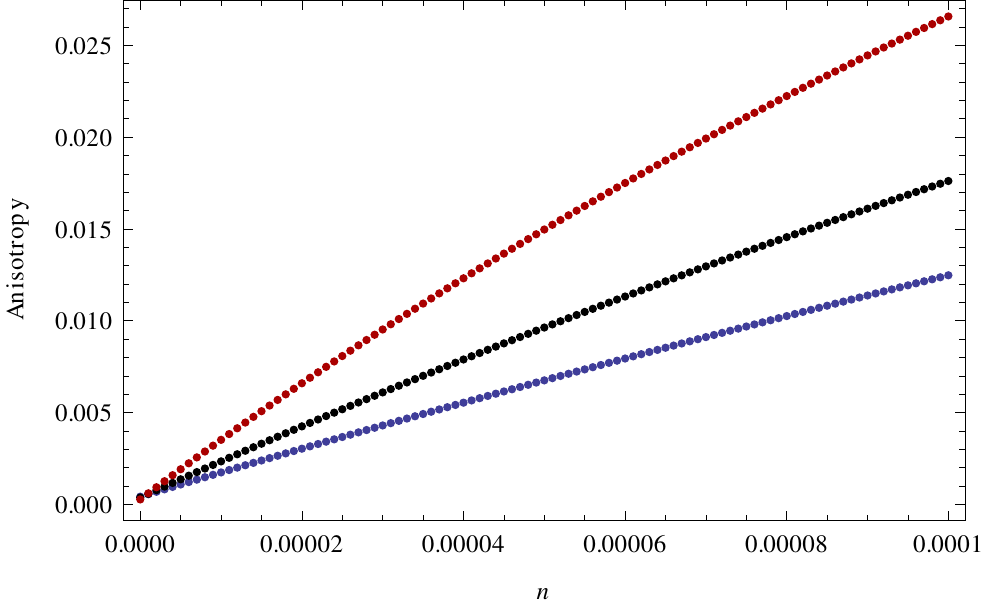}
\caption{Anisotropy vs $n$ plot. The plot is obtained by slowly increasing $n$ 
from $n=0$ to $n=10^{-4}$ for $\rho=40$ (Blue dots), $\rho=60$ (Black dots), $\rho=100$ (Red dots). }
 \end{figure}
 
\noindent The slow-roll parameters are given by,
\begin{equation}
 \epsilon_{H} =-\displaystyle \frac{\dot{H}}{H^2}=\frac{1}{\zeta},\qquad  \delta_{H}= \dot{\xi} H = 2 n\zeta 
\end{equation}
Since $n>0$ and $\zeta>0$, $\delta_H$ being proportional to $n$ is positive in our case, so
(\ref{expr_c}) indicates that the anisotropy increases when Gauss-Bonnet correction is included. 
Then an analogous relation of (\ref{expr_c}) in case of power-law anisotropic solutions can be expressed as 
\begin{equation}
 \frac{\eta}{\zeta}=\frac{c-1}{3 c}(\epsilon_H+\delta_H) \label{temp},
\end{equation}
which implies,
\begin{equation}
 c= \frac{1+2n\zeta^2}{1+2n\zeta^2-3\eta}.
\end{equation}
Now using (\ref{zeta}) and (\ref{eta}) we obtain,
\begin{equation}
 c=\displaystyle\frac{1+\frac{2 n \left[\lambda ^3 (3 n+1)+4 \lambda ^2 (5 n+2) \rho +
 4 \lambda  \left((11 n+3) \rho ^2+2\right)+32 n \rho ^3\right]^2}{9 \lambda^2 (\lambda +
 2 \rho )^2 (\lambda  (3 n+2)+6 n \rho )^2}}{1+\frac{2 n \left[\lambda ^3 (3 n+1)+
 4 \lambda ^2 (5 n+2) \rho +4 \lambda  \left((11n+3) \rho ^2+
 2\right)+32 n \rho ^3\right]^2}{9 \lambda ^2 (\lambda +2 \rho )^2 (\lambda  (3 n+2)+
 6 n \rho )^2}-\frac{\lambda ^3 (3 n+4)+2\lambda ^2 (7 n+4) \rho +
 4 \lambda  \left(5 n \rho ^2-4\right)+8 n \rho ^3}{2 \lambda  (\lambda +2 \rho ) (\lambda  (3 n+2)+6 n \rho )}}.
\end{equation}
For $\rho>>1$, $\lambda<<1$ and $n<<1$, $c$ is always positive and greater than $1$.
With $c>1$, the vector field grows during slow roll inflation which is a requirement for 
anisotropic effects to be non-zero in presence of Gauss-Bonnet corrections.
We note here above relation for 
$c$ reduces to same expression obtained in \cite{Kanno} for non-Gauss-Bonnet case.\\ 
In the following figure, the anisotropy $\Sigma/H$ is plotted against $\alpha$ for given values of 
$\xi_0, V_0, f_0 \lambda$ and $c$. 
The quantity $\Sigma/H$ decreases with $\alpha$. 
The nearly constant region ( of the order of $10^{-14}$) in the plot corresponds to the isotropic phase. 
The universe then passes through an anisotropic 
inflationary phase when anisotropy becomes maximum and finally anisotropy decays down.
The plot is obtained numerically using (\ref{ttcomp})-(\ref{scalar_comp}) with $\kappa=1$ and $c=2$.
The boundary condition for $\dot{\alpha}$ is determined by solving the scale factor equation 
with initial conditions : $\alpha(0)=0, \phi(0)=10^{-5},\sigma(0)=0,\dot{\phi}(0)=10^{-25},
\dot{\sigma}(0)=10^{-20}, v(0)=0, \dot{v}(0)=2.4 \times 10^{-100}$ and
$\xi_0=10^{-10}$, $V_0=10^{-10}$,$f_0=0.1$, $\lambda=10^{-3}$.
\begin{figure}[h]
 \begin{minipage}[b]{0.5\textwidth}
 \includegraphics[width=2.7 in]{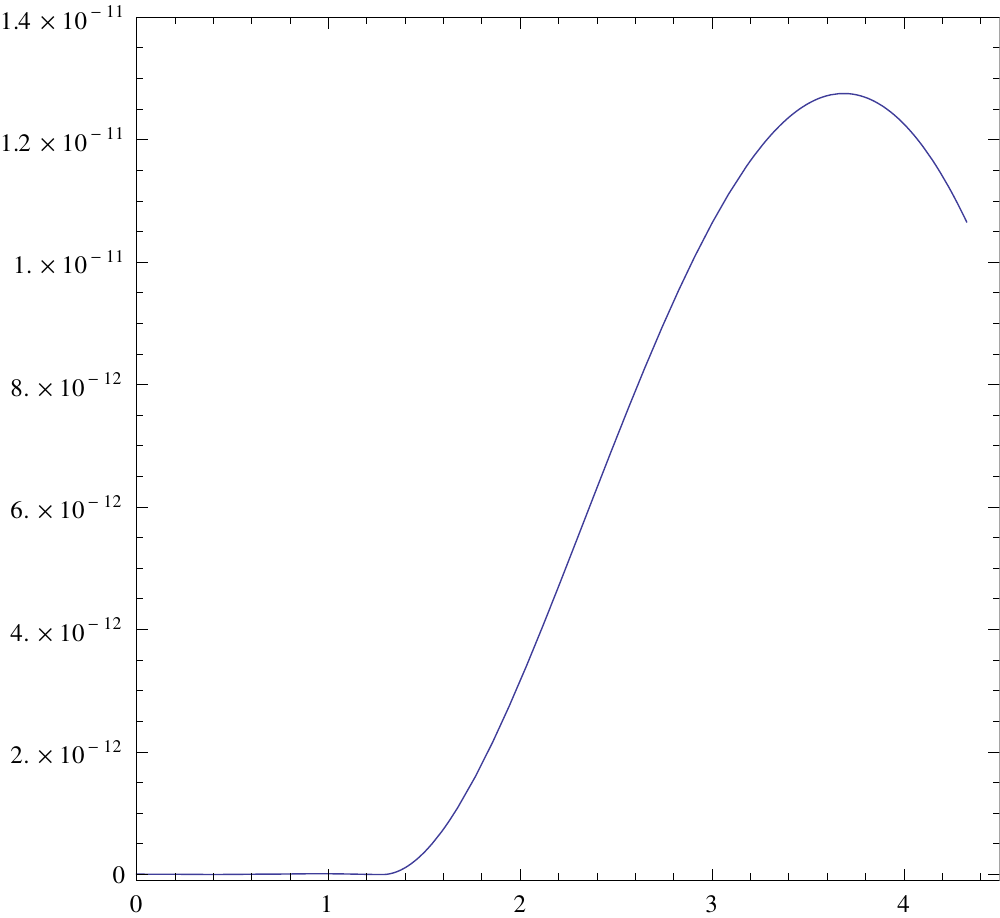}
 \caption{Plot of $\frac{\Sigma}{H}$ vs $\alpha$}
 \end{minipage}
 \end{figure}

\section{Stability analysis of anisotropic power-law solutions}
We will now determine fixed phase points, then stability of inflationary solutions 
will be examined around these fixed points. 
Here, the e-folding number $\alpha$ is taken as the 
time co-ordinate so that $d\alpha=\dot{\alpha}dt$.
The equations of motion can be 
expressed in terms of dimensionless quantities which are defined as follows
\begin{equation}
 X=\displaystyle\frac{\dot{\sigma}}{\dot{\alpha}},\quad Y=\kappa \displaystyle\frac{\dot{\phi}}{\dot{\alpha}},\quad
 Z=\kappa f(\phi)e^{-\alpha+2\sigma}\frac{\dot{v}}{\dot{\alpha}}.
\end{equation}
Using (\ref{v}), (\ref{ttcomp}) is expressed as,
\begin{equation}
-\kappa^2\displaystyle\frac{V(\phi)}{\dot{\alpha}^2}=3(X^2-1)+\frac{1}{2}Y^2+\frac{1}{2}Z^2-3 N_1(1-2X)(1+X)^2.
\end{equation}
Since we assume a positive inflaton potential i,e. $V(\phi)>0$ therefore we have
\begin{equation}
3(X^2-1)+\frac{1}{2}Y^2+\frac{1}{2}Z^2-3 N_1(1-2X)(1+X)^2 < 0,   \label{d_0}
\end{equation}
where $N_1=\kappa^2 \dot{\xi} \dot{\alpha}$ is dimensionless. 
Using (\ref{v}), the equations of motion (\ref{aniso-comp})-(\ref{scalar_comp}) can be recast in terms
of dimensionless variables as
\begin{eqnarray}
 \displaystyle\frac{dX}{d\alpha}&=&\displaystyle\frac{1}{3}Z^2(X+1)+X\left[3(X^2-1)+\frac{1}{2}Y^2\right] \nonumber\\[1mm]
&& - N_1\left[X\left(\frac{7}{2}+ 3X - \frac{9}{2}X^2 + 4X^3\right)-(1+X)^2\frac{\eta}{\zeta^2}\right]-
\frac{1}{2} N_2 X (1+X)^2,         \label{d_1} \\[2mm] 
 \displaystyle\frac{dY}{d\alpha}&=& (Y+\lambda)\left[3(X^2-1)+\frac{1}{2}Y^2\right]+\frac{1}{3}Y Z^2+ 
 \left(\rho+\frac{\lambda}{2}\right)Z^2 \nonumber \\[1mm]
&&+\frac{3\lambda}{2} N_1 (1+X)\left[(-1+X+ 4X^2)-\zeta (1-X-2X^2)-6\frac{\eta}{\zeta}X\right]
 \nonumber \\
&&-N_1 Y\left(\frac{1}{2}-\frac{9}{2}X^2-4X^3 +\frac{1}{\zeta}-
X \frac{\eta}{\zeta^2}\right)-\frac{1}{2} N_2 Y(-1+X^2),   \label{d_2}\\[2mm] 
 \displaystyle\frac{dZ}{d\alpha}&=& Z\left[3(X^2-1)+\frac{1}{2}Y^2-\rho Y +1-2X+\frac{1}{3}Z^2 \right] \nonumber \\[1mm]
&&- N_1 Z\left(\frac{1}{2}-\frac{9}{2}X^2 - 4X^3 + \frac{1}{\zeta}- X\frac{\eta}{\zeta^2}\right)
-\frac{1}{2}N_2 Z(-1+X^2)     \label{d_3}
\end{eqnarray}
 and we define $N_2=\kappa^2 \ddot{\xi}$. 
 For Gauss-Bonnet power-law solutions, using (\ref{alpha}), (\ref{phi}), we get $N_1=2 n \zeta$ and $N_2=2n$.
Therefore, in absence of Gauss-Bonnet correction, $N_1$ and $N_2$ vanish and (\ref{aniso-comp})-(\ref{scalar_comp})
reduce to those for non-Gauss-Bonnet case \cite{Kanno}.
\subsection{Determination of phase points}
The fixed phase point corresponds to that point which does not evolve with time. Therefore in the given set-up
fixed points in the phase space are determined by solving 
\begin{equation}
\displaystyle\frac{dX}{d\alpha}=0, \qquad \displaystyle\frac{dY}{d\alpha}=0,
\qquad \displaystyle\frac{dZ}{d\alpha}=0 \label{phase_con}.
\end{equation}
\subsubsection{Isotropic phase points}
Since isotropy implies $\eta=0$ therefore we have $X=0$. 
Substitution of $X=0$ and $N_1=N_2 \zeta$ in (\ref{d_1}) leads to $Z=0$. 
Now with $X=0=Z$, $N_1=N_2 \zeta = 2 n \zeta$ and using conditions (\ref{phase_con}), (\ref{d_2}) yields 
\begin{equation}
\frac{1}{2} \left(-6 \lambda +Y^3+\lambda  Y^2-6 Y\right)+ n \left(-3 \zeta ^2 \lambda -3 \zeta  \lambda -
 \zeta  Y-Y\right)=0  \label{solY}.
\end{equation}
It is a cubic equation which gives three roots of $Y$. 
Since the inflaton potential $V(\phi)>0$,
therefore the root of $Y$ which satisfies positivity of $V(\phi)$ is taken and other two roots are discarded.
The particular root of $Y$ satisfying the condition (\ref{d_0}) is given by
\begin{equation}
Y=-\displaystyle \frac{\lambda}{3}-\frac{2^{1/3}\,p_1}{3 \left(p_2+\sqrt{4p_1^3+p_2^2}\right)^{1/3}},  \label{isoY}
\end{equation}
where $p_1=-\lambda ^2-6 \zeta  n-6 n-18$ and
$p_2=2 \lambda (-\lambda ^2+ 54 + 81 \zeta ^2  n+ 72 \zeta  n-9  n)$.\\
The complete solution of $Y$
is obtained by substituting $\zeta$ i,e. (\ref{iso_zeta}) in (\ref{isoY}).
Now, $Y$ can be expressed in the leading order of $n$ plus higher order terms in $n$ as
\begin{equation}
Y= -\lambda- \frac{2(2+\lambda^2)}{\lambda^3}n +...  \label{Yiso}   
\end{equation}
The phase point $Y$ exactly reduces to
$Y=-\lambda$ when $n=0$ which is the non Gauss-Bonnet counterpart of $Y$ obtained in \cite{Kanno}.
We mention here that other two roots of $Y$ by solving (\ref{solY}) reproduces $Y=\pm \sqrt{6}$
in $n\rightarrow 0$ limit but we will not consider them here as these 
solutions do not satisfy (\ref{d_0}).
So the isotropic fixed point is given by $(0,Y,0)$.\\
It may be seen that this isotropic fixed phase point corresponds to isotropic power-law inflation 
given by (\ref{iso_cons0}) and (\ref{iso_zeta}).    
In Figure. 2 we have plotted the variation of $Y$ with $\lambda$ for four different
values of $n$ to compare the deviations in the phase point $Y$ due to Gauss-Bonnet corrections.
The deviations are small but particularly appreciable for lower $\lambda$. 
\begin{figure}[h]
\centering
\includegraphics{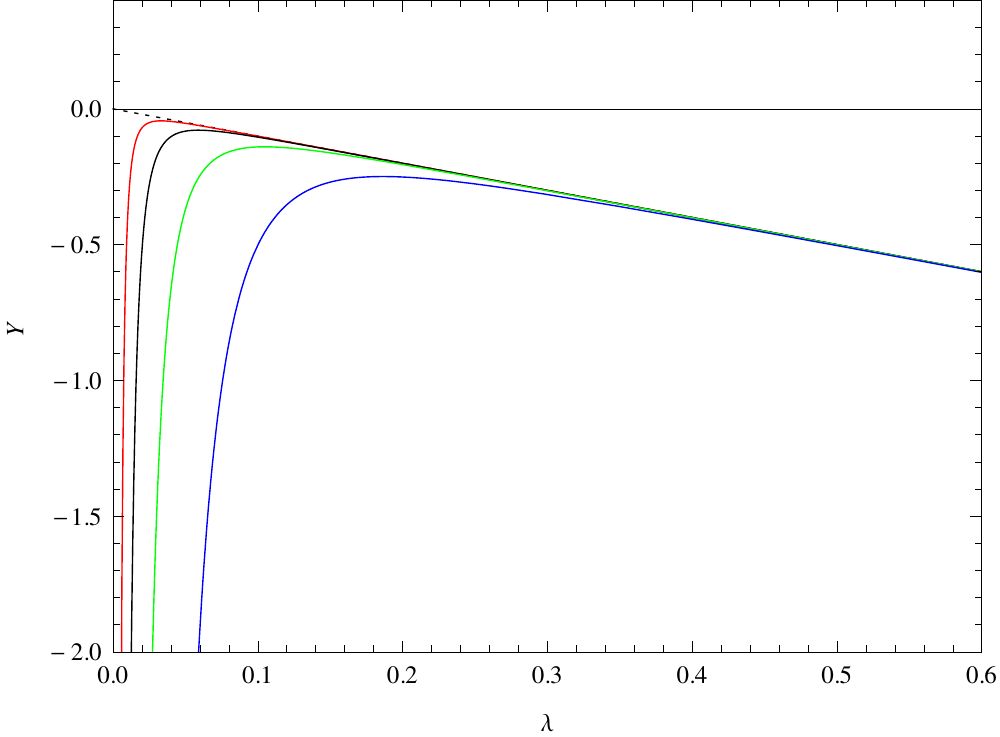} 
\caption{Plot of $Y$ vs $\lambda$ in the isotropic power-law solution. 
Dashed line ($n=0$), red line ($n=10^{-7}$),
black line ({$n=10^{-6}$}), green line ({$n=10^{-5}$}), blue line ({$n=10^{-4}$}) describe plots for 
different values of n.}
\end{figure}
\subsubsection{Anisotropic phase points}
We now determine the anisotropic phase points using equations
(\ref{d_1})-(\ref{d_3}) and  (\ref{phase_con}) where
we now have $\eta \neq 0$. From (\ref{d_1})
\begin{eqnarray}
 Z^2 &=& -\frac{3X}{(X+1)}\left[3(X^2-1)+\frac{1}{2}Y^2\right]+\frac{6 n \zeta}{(X+1)}\left[X\left(\frac{7}{2}+ 
 3X - \frac{9}{2}X^2 + 4X^3\right)-(1+X)^2\frac{\eta}{\zeta^2}\right] \nonumber\\[2mm]
 &&+3n X (X+1)    \label{Z1}
\end{eqnarray}
and similarly from (\ref{d_3}), we can write
\begin{equation}
 Z^2=-\frac{3 \left(-4 \zeta -2 \zeta ^2 n-2 \zeta  n+16 \zeta ^2 n X^3+18 \zeta ^2 n X^2-2 \zeta  n X^2+4 \eta  n X+6 \zeta  X^2-4 \zeta  X+\zeta 
   Y^2-2 \zeta  \rho  Y\right)}{2 \zeta }.  \label{Z2}
\end{equation}
Subtracting (\ref{Z2}) from (\ref{Z1}), we obtain
\begin{equation}
\begin{array}{rcl}
\zeta  \left[2 n \left(X^2-1\right)+2 X^2-2 \rho  (X+1) Y-2 X+Y^2-4\right]\\[1mm]
 +2 n (X+1)^2 (8 X-1)\zeta ^2
 - 4 n (X+1)\eta & = & 0,
\end{array}
 \end{equation}
which is a quadratic equation in $Y$ and on solving gives two roots of $Y$ in terms of $X$ as
\begin{equation}
\begin{array}{rcl}
 Y_{(-,+)} & &=  (1+X) \rho \\[1mm]
&& \mp \, \displaystyle \frac{\sqrt{-\zeta  (1+X) \left[-4 \eta  n+2 \zeta ^2 n \left(8 X^2+7 X-1\right)+
 \zeta  \left(-4+2 n (X-1)-\rho ^2(1+X)+2 X\right)\right]}}{\zeta }.  \nonumber
\end{array}
 \end{equation}
Out of the two solutions, we shall take the first solution of $Y$ i,e. $Y_{(-)}$ because it reduces to
the non-Gauss-Bonnet fixed phase point in $n \rightarrow 0$ limit. Removing the subscript, we obtain $Y$ 
in terms $X$ as
\begin{equation}
\begin{array}{rcl}
 Y  & &= (1+X) \rho \\[1mm]
\, &&  -\, \displaystyle\frac{\sqrt{-\zeta  (1+X) \left[-4 \eta  n+2 \zeta ^2 n \left(8 X^2+7 X-1\right)+
 \zeta  \left(-4+2 n (X-1)-\rho ^2(1+X)+2 X\right)\right]}}{\zeta}.    \label{aniY}
\end{array} 
 \end{equation}
When we substitute (\ref{aniY}) in either one of two equations (\ref{Z1}) or (\ref{Z2}), we get $Z^2$ in terms $X$,
\begin{equation}
\begin{array}{rcl}
 Z^2 & = & \displaystyle\frac{3\left[\zeta  X^2 \left(2n(3 \zeta +1) -\rho ^2-2\right)-2 \eta  n \right]}{\zeta}
+ \displaystyle\frac{3X\left(-\zeta  \rho ^2+\zeta +6 \zeta ^2 n-4 \eta  n \right)}{\zeta} \\[2mm]
&& + \displaystyle\frac{3 X \rho\sqrt{\zeta  (X+1) \left[4 \eta  n-2 \zeta  
(n (\zeta  (X+1) (8 X-1)+X-1)+X-2)+\zeta  \rho ^2 (X+1)\right]}}{\zeta}.   \label{aniZ}
\end{array}
 \end{equation}
Now, substituting $Y$ and $Z^2$ from (\ref{aniY}) and (\ref{aniZ}) in (\ref{d_2}) and from the condition 
$\displaystyle \frac{dY}{d\alpha} =0 $, we get a polynomial equation of $X$ as 
\begin{equation}
\begin{array}{lcr}
6 n \,\zeta ^3 \,\lambda  (X+1)^2 (2 X-1)-2 n \,\eta (4 X+1) \,(\lambda +2 \rho )\\[1mm]
+\left[2-2 \lambda  \rho -4 \rho ^2+X \left(\lambda  \rho +2 \rho ^2-4\right)\right]a(X)\\[1mm]
+2n \zeta ^2 (X+1) 
\left[\lambda  \left(4 X^2+5 X-2\right)+2 \rho  \left(-8 X^2+2 X+1\right)\right]\\[1mm]
+\zeta \left[
2 \rho \left(n 
\left(4 X^2+2\right)+2 \rho ^2-\left(\rho ^2+6\right) X^2+\left(\rho ^2+6\right) X+3
\right)\right] \\[1mm]+ 
\zeta \left[\lambda \left(n \left((4-36 \eta ) X^2-36 \eta  X+2\right)-(X-2) \left(\rho ^2+
\left(\rho ^2+2\right) X-1\right)\right)
\right] & = & 0,            \label{aniX}
\end{array}
\end{equation}
where we define,\\
$a(X)=\sqrt{\zeta  (X+1) \left[4 \eta  n-2 \zeta ^2 n \left(8 X^2+7 X-1\right)+\zeta  
\left(-2 n (X-1)+\rho ^2+\left(\rho ^2-2\right)
   X+4\right)\right]}.\\[2mm]$
After substituting (\ref{zeta}) and (\ref{eta}), the above equation can be solved for
determining possible real roots of $X$.
Here we will solve (\ref{aniX}) numerically.\\ 
Since anisotropic power-law inflation requires $n<<1$, $\lambda<<1$ and $\rho>>1$, the roots of $X$ can be computed
for given values of $n$, $\lambda$, $\rho$ satisfying these conditions. 
Then using roots of $X$ determined from (\ref{aniX}) in (\ref{aniY}) and (\ref{aniZ}) $Y$  and $Z^2$ are evaluated.
Here, we have taken $\rho=40$ and $\lambda=0.1$ for our numerical calculations. 
Figure 3(a) is the plot of (\ref{aniX}) vs $X$ in which the point of intersection on the
X-axis for a particular value of $n$ gives the root of $X$ for that value of $n$ 
which can be seen from Figure 3a.
When $n=0$, for $\rho=40$ and $\lambda=0.1$, (\ref{aniX}) $X=4.1 \times 10^{-4}$.
Slowly the Gauss-Bonnet correction is turned on and $n$ is increased from zero. 
It is found that deviation of $X$ occurs from $n=0$ to $n=10^{-5}$ where
$X$ increases by an order of $10$ as compared to its value at $n=0$. 
We mention here that the parameter $n$ due to Gauss-Bonnet correction is restricted up to the value of $10^{-5}$
(where $X=17.3 \times 10^{-4}$) beyond which $X$ shifts completely from
$n=0$ result so that Gauss-Bonnet gravity cannot be treated as correction.
 \begin{figure}[h]
 \begin{minipage}[b]{0.5\textwidth}
 \includegraphics[width=3.3 in]{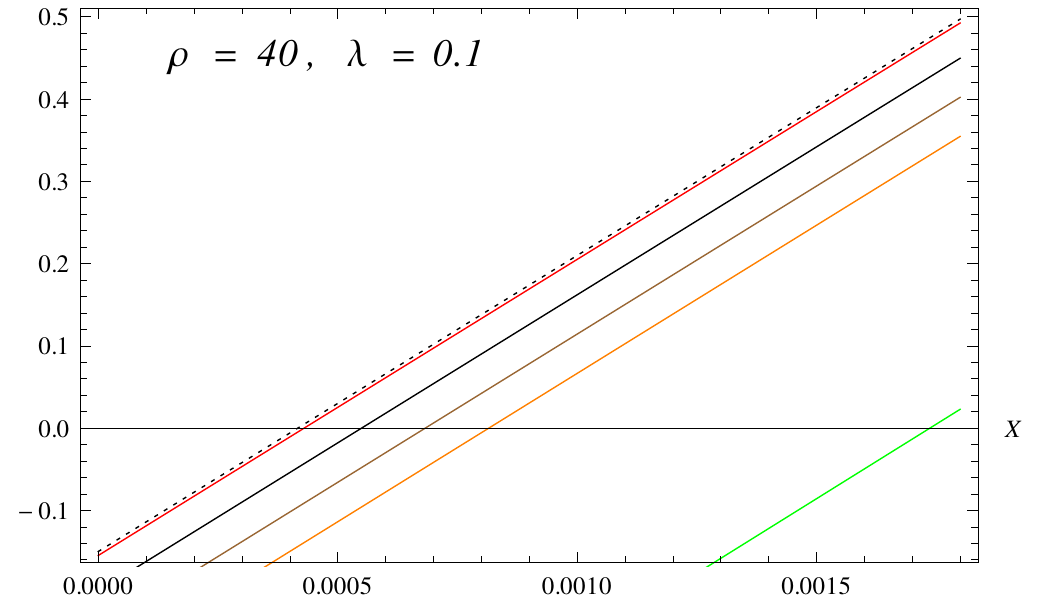}
 \caption*{(a) Roots of (\ref{aniX}) for different $n$}
 \end{minipage}
   \hfill
   \begin{minipage}[b]{0.5\textwidth}
     \includegraphics[width=3.2 in]{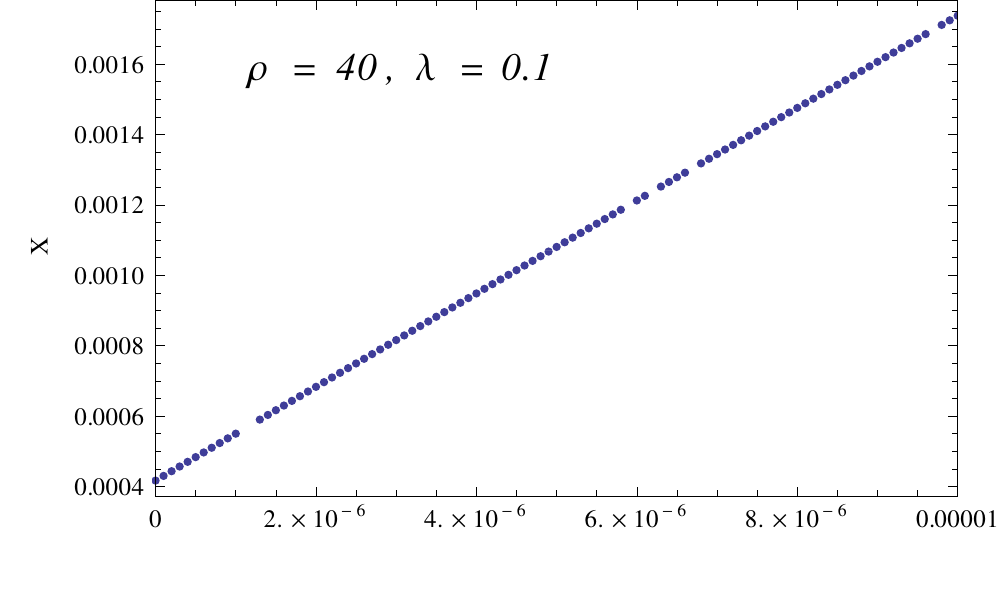}
      \caption*{(b) Plot of $X$ vs $n$}
   \end{minipage}
    \caption{In figure (a) location of $X$ with different n are shown. Dashed line : $n=0$,$X=4.1\times 10^{-4}\,$,
  red line: $n=10^{-7}$, $X=4.3\times 10^{-4}\,$, Black line: $n=10^{-6}$, $X=5.5\times 10^{-4}\,$,
 Brown line: $n=2 \times 10^{-6}$, $X=6.8\times 10^{-4}\,$, Orange line: $n=3 \times 10^{-6}$, $X=8.1\times 10^{-4}\,$,
 Green line: $n=10^{-5}$, $X=17.3\times 10^{-4}.$  Figure (b), 
 the plot shows the shift of the phase point $X$ as $n$ is increased from $n=0$. }
 \end{figure}
The phase points $X$ as plotted in Figure 3(b) are obtained as roots of (\ref{aniX}). 
Using these values of $X$, 
$Y$ and $Z$ can be determined using (\ref{aniY}) and (\ref{aniZ}) respectively. 
These values of anisotropic fixed phase points $Y$ and $Z^2$ for different 
values of $n$  with $\rho=40$ and $\lambda=0.1$ are shown in Figure 4 and Figure 5 respectively.
These plots present the corresponding deviations of $Y$ and $Z^2$ from $n=0$ results.
We mention here that relations of $X,Y$ and $Z^2$ reduce to non- Gauss-Bonnet results obtained in \cite{Kanno}. 
\begin{figure}[h]
 \begin{minipage}[b]{0.5\textwidth}
 \includegraphics[width=3.0 in]{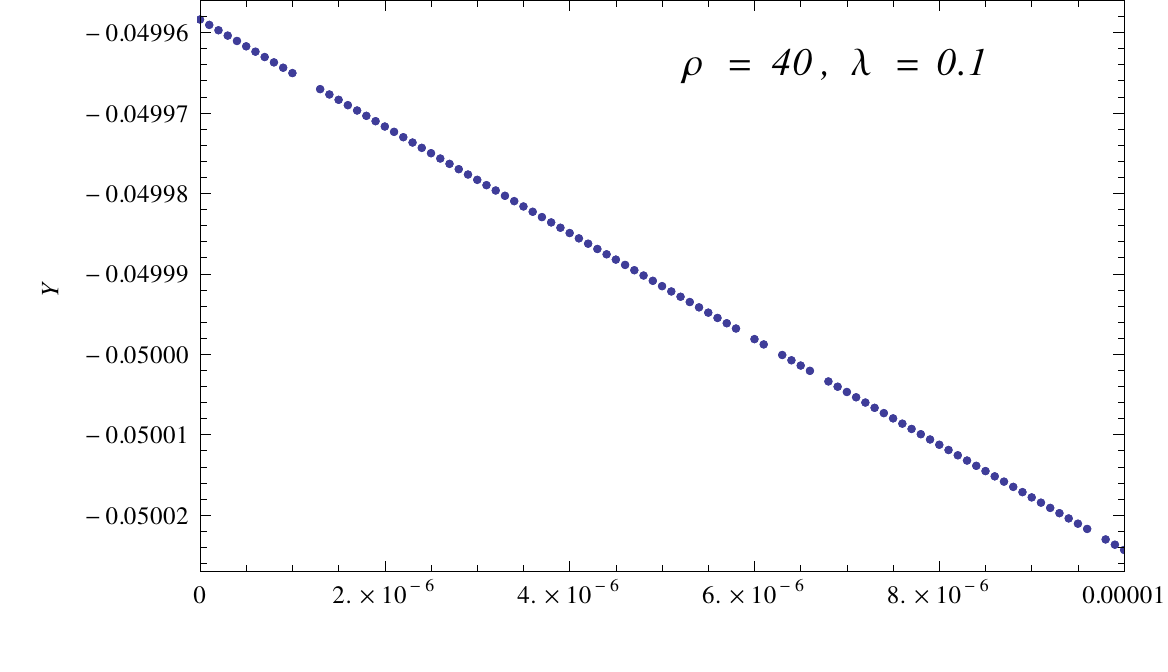}
 \caption{Plot of $Y$ vs $n$ }
 \end{minipage}
   \hfill
   \begin{minipage}[b]{0.5\textwidth}
     \includegraphics[width=3.0in]{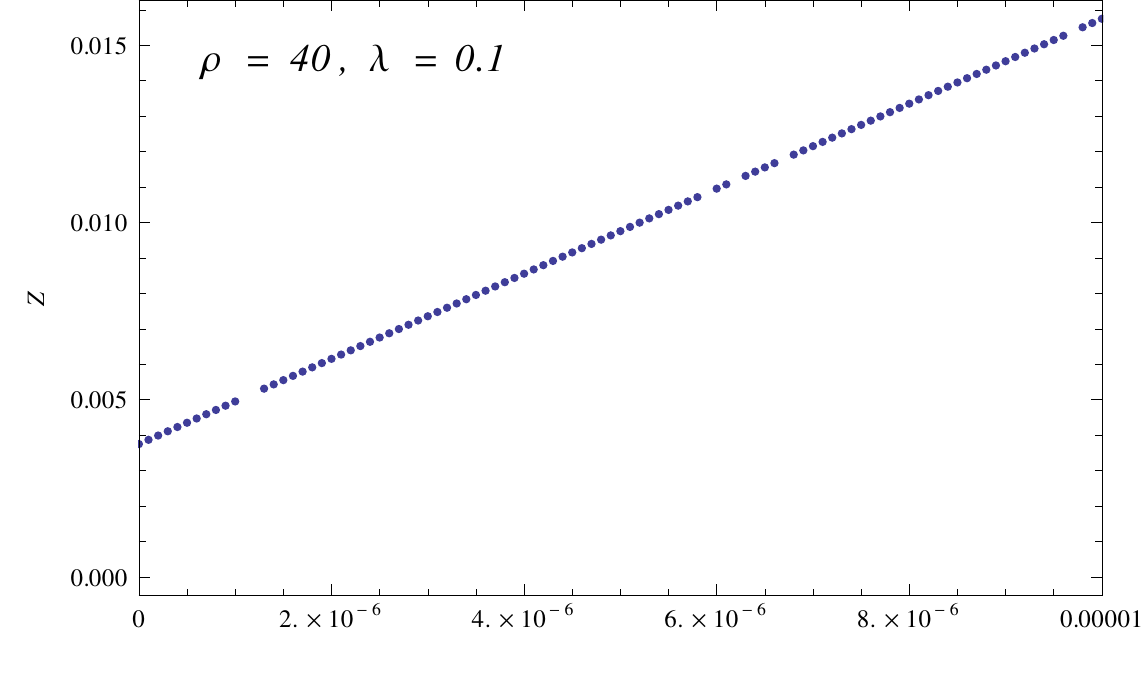}
 \caption{Plot of $Z$ vs $n$}
 \end{minipage}
 \end{figure}
 \pagebreak

\subsection{Stability analysis }
We now investigate the linear stability of isotropic and anisotropic fixed phase points in order 
to investigate the impact of Gauss-Bonnet term on the stability
of the obtained solutions.
The linearized equations necessary for the stability analysis are determined from
(\ref{d_1})-(\ref{d_3}) and are given by
\begin{equation}
\begin{array}{rcl}
\displaystyle\frac{d\delta X}{d \alpha} &=&  \left(9 X^2+\displaystyle\frac{Y^2}{2}+
\displaystyle\frac{Z^2}{3}-3\right) \delta X + 
 X Y \delta Y +\displaystyle\frac{2}{3} (X+1) Z \delta Z  \label{L1} \\[3mm]
&& -2 \zeta  n \left[16 X^3-\displaystyle\frac{27 X^2}{2}+\displaystyle\frac{2 \eta  (X+1)}{\zeta ^2}+6 X+
\displaystyle\frac{7}{2}\right]\delta X -
n  \left[(X+1)^2+2 X (X+1)\right]\delta X ,
\end{array}
\end{equation}

\begin{eqnarray}
  \frac{d\delta Y}{d \alpha} &=& 6 X (\lambda +Y)\delta X
  +3\left[ \left(X^2-1\right)+\frac{Y^2}{2}+Y (\lambda +Y)+\frac{Z^2}{3}\right]\delta Y+ Z \left(\lambda +2 \rho +\frac{2 Y}{3}\right)
  \delta Z  \nonumber \\[1mm]
  && +2 n X \left[9 \zeta ^2 \lambda +15 \zeta  \lambda -18 \eta  \lambda +9 \zeta ^2 \lambda  X+18 \zeta  \lambda  X+12 \zeta  X Y+(9 \zeta -1)
   Y\right]\delta X \nonumber \\[1mm]
   && + 2 n \left(\frac{\eta  Y}{\zeta }-9 \eta  \lambda \right) \delta X+
   n \left[-\zeta +8 \zeta  X^3+(9 \zeta -1) X^2+\frac{2 \eta  X}{\zeta }-1\right]\delta Y , \label{L2} \\[2mm]
   \frac{d\delta Z}{d \alpha} & = & 2 Z (3 X-1) \delta X +  Z (Y-\rho ) \delta Y+
    \left[3 \left(X^2-1\right)-2 X+\frac{Y^2}{2}-\rho  Y+Z^2+1\right]\delta Z \nonumber \\[1mm]
    && + 2 n Z \left[\frac{\eta }{\zeta }+12 \zeta  X^2+(9 \zeta -1) X\right]\delta X
    + n \left[-\zeta +8 \zeta  X^3+(9 \zeta -1) X^2+\frac{2 \eta  X}{\zeta }-1\right]\delta Z.  \label{L3} \nonumber
\\    
\end{eqnarray}
\subsubsection{Isotropic case}
The isotropic inflation corresponds to $\eta = 0$ and the corresponding fixed phase point is $(0,Y,0)$. 
Substituting $\zeta$ and $Y$ from (\ref{iso_zeta}) and (\ref{isoY}) 
in (\ref{L1})-({\ref{L3}}) and in the leading order of $n$ these linearized equations are given by 
\begin{equation}
\begin{array}{rcl}
\displaystyle\frac{d\delta X}{d \alpha} & =&  \left[\left(\displaystyle\frac{\lambda ^2}{2}-3\right)+
\left(1-\displaystyle\frac{10}{\lambda ^2}\right) n \right]\delta X, \\[3mm]
\displaystyle \frac{d\delta Y}{d \alpha} & = & \left[\left(\displaystyle \frac{\lambda ^2}{2}-3\right)+
\left(\displaystyle\frac{6}{\lambda ^2}+3\right) n\right]\delta Y, \\[3mm]
\displaystyle \frac{d\delta Z}{d \alpha} &= &\left[\displaystyle \left(-2+\frac{\lambda^2}{2}+\lambda \rho\right)+
\frac{(2+\lambda^2)(\lambda+2\rho)}{\lambda^3}n\right]\delta Z.
\end{array}
 \end{equation}
For stable isotropic solutions, we must have $\displaystyle \frac{d\delta X}{d \alpha}<0$, 
$\displaystyle \frac{d\delta Y}{d \alpha}<0$ and $\displaystyle \frac{d\delta Z}{d \alpha}<0$ which 
suggests $\lambda <<1$, $n<<1$ and in addition the
following condition
\begin{equation}
(\lambda^2+2\lambda\rho) \left[1+\displaystyle\frac{2 n}{\lambda^4} (2+\lambda^2)\right]<4 \label{rel}
\end{equation}
is required to be satisfied. 
Thus for isotropic stable power-law inflation, $\rho$,$\lambda$ and $n$ are related by (\ref{rel}).
Therefore if this condition is not satisfied, the isotropic inflation becomes
unstable.
Since $n<<1$, we find that (\ref{rel}) does not remain valid when $\rho>>1$. 
This is in fact a condition for anisotropic inflation. 
So, (\ref{rel}) suggests that for stable anisotropic inflationary solutions $\rho>>1$ in addition to 
the conditions $\lambda<<1$ and $n<<1$.

\subsubsection{Anisotropic case}
In order to examine stability of anisotropic inflationary solutions, we again concentrate on
linearized equations. 
Substituting $\zeta$ and $\eta$ from (\ref{zeta}) and (\ref{eta}) in (\ref{L1})-({\ref{L3}}) and
in the leading order in $n$, we obtain
\begin{eqnarray}
 \displaystyle \frac{d\delta X}{d \alpha} & = & \left(9 X^2+\frac{Y^2}{2}+\frac{Z^2}{3}-3\right)\delta X
 + X Y\delta Y +\frac{2}{3} (X+1) Z \delta Z - n\left(3  X^2 +4  X+1\right)\delta X \nonumber \\[3mm]
&-& n\frac{ \left(\lambda ^2+8 \lambda  \rho +12 \rho ^2+8\right) \left(16 X^3-\displaystyle\frac{27 X^2}{2}+
\displaystyle\frac{96 \lambda  (X+1) (\lambda +2 \rho )
   \left(\lambda ^2+2 \lambda  \rho -4\right)}{\left(\lambda ^2+8 \lambda  \rho +12 \rho ^2+8\right)^2}+6 X+\frac{7}{2}\right)}{6 \lambda 
   (\lambda +2 \rho )} \delta X, \nonumber   
   \\ [1mm]  \label{L1-aniso}
   \end{eqnarray}
   \begin{equation}
    \begin{array}{rcl}
\displaystyle \frac{d\delta Y}{d \alpha} & = & 
\displaystyle\frac{1}{6} \left[36 X (\lambda +Y)\delta X +  \left(18 X^2+9 Y^2+6 \lambda  Y+2 Z^2-18\right)\delta Y+
2  Z (3 \lambda +6 \rho +2 Y)\delta Z\right]   \\[2mm]
 &+ & n  \Bigg[\displaystyle\frac{\left(\lambda ^2 (25 (X-3) X-48)+
 4 \lambda  X (4 X-1) Y+48 X (X+1)\right)\delta X}{8 \lambda}\\[2mm]
 &&  \qquad \,+\displaystyle\frac{ \lambda \left(X^2 (8 X-3)-13\right) \delta Y}{12 \lambda } \Bigg]\\[2mm]
  &+& n\left[ \displaystyle\frac{9 \rho ^2 X (X+1)\delta X}{2 \lambda }+\frac{8 X (X+1)\delta X}{\lambda  (\lambda +2 \rho )^2}
  + \frac{8 \left(\lambda ^2+2 \lambda  \rho -4\right) ( Y\delta X+  X\delta Y)}{(\lambda +2 \rho ) (\lambda +
  6 \rho )+8}\right]\\[2mm]
  & + & n \left[\displaystyle\frac{3 \rho X (\lambda  (13 X+11)+8 X Y+6 Y)\delta X+ \rho (X+1) \left(8 X^2+X-1\right)
  \delta Y}{2 \lambda }\right] \\[2mm]
  &+& n \left[\displaystyle\frac{ 12 (\lambda  (5 X^2+ 16X +6)+X (4 X+3) Y)\delta X+ 2 (X+1) \left(8 X^2+X-1\right)\delta Y}
  {3 \lambda  (\lambda +2 \rho )}\right],
 \label{Lin2} 
 \end{array}
  \end{equation}
 
 \begin{equation}
   \begin{array}{rcl}
   \displaystyle \frac{d\delta Z}{d \alpha}& =& Z (6 X-2)\delta X+ Z (Y-\rho )\delta Y+
    \left(3 X^2-2 X+\frac{Y^2}{2}-\rho  Y+Z^2-2\right)\delta Z \\[2mm]
     &+& \displaystyle \frac{1}{12}n \Bigg[ \frac{6 \rho  \left(6 X (4 X+3) Z \delta X + 
     (X+1) \left(8 X^2+X-1\right)\delta Z \right)}{\lambda } \\[1mm]
     && \qquad \,+ \displaystyle\frac{48  X (4 X+3) Z \delta X + 8 (X+1)
   \left(8 X^2+X-1\right)\delta Z}{\lambda  (\lambda +2 \rho )}\\[1mm]
   && \qquad \,\, +\displaystyle \frac{96 \left(\lambda ^2+2 \lambda  \rho -4\right) (Z\delta X +  X\delta Z)}
   {(\lambda +2 \rho) (\lambda +6 \rho )+8}
  + 6 X Z (4X-1)\delta X + (8X^3-3X^2-13)\delta Z \Bigg].    \label{Lin3}
   \end{array}
   \end{equation}
Now stability of anisotropic solutions are studied from (\ref{L1-aniso}),(\ref{Lin2})
and (\ref{Lin3}).
Since Gauss-Bonnet gravity describes higher
order corrections so leading order terms in $n$ give most dominant 
contribution compared to sub-leading orders which we have also observed earlier from our study.
As $\rho>>1$ and $\lambda<<1$ are required conditions for inflation, we have determined phase points
for $\rho=40, \, \lambda=0.1$ in the range $n=0$ to $n=10^{-5}$ 
using (\ref{aniX}), (\ref{aniY}) and (\ref{aniZ}) respectively.
The evolution of the phase points with $n$ are then plotted in Fig. 3(b), Fig. 4 and Fig. 5 respectively. 
Using these phase points, we shall investigate the stability of anisotropic power-law solution.
For our analysis, we again take $\rho=40$ and $\lambda=0.1$.
We will consider different values of $n$ in the increasing
order of magnitude and solve (\ref{L1-aniso}),(\ref{Lin2})
and (\ref{Lin3}). 
\begin{itemize}
 \item For $n=10^{-7}$, approximately linear equations can be expressed as 
\begin{equation}
 \displaystyle \frac{\delta X}{d \alpha} = -2.9988\, \delta X     \label{sta1a} \\
\end{equation}
\begin{equation}
 \displaystyle \frac{d\delta Y}{d \alpha} = -3.0012 \,\delta Y + 0.6195\, \delta Z     \label{sta1b}
\end{equation}
\begin{equation} 
\displaystyle \frac{d\delta Z}{d \alpha}  =  -0.1521\, \delta  Y     \label{sta1c}
\end{equation}
These linear equations can be solved by setting, 
\begin{equation}
\delta X= e^{\Omega \alpha} \delta \tilde{X},\quad \delta Y= e^{\Omega \alpha} \delta \tilde{Y},\quad 
\delta Z= e^{\Omega \alpha} \delta \tilde{Z}     \label{test_stab}
\end{equation}
From (\ref{sta1a}), we get $ \Omega= -2.998$. 
Solving (\ref{sta1b}) and (\ref{sta1c}) give $\Omega= -2.969, -0.031$ which suggest anisotropic inflation
solutions with Gauss-Bonnet corrections are stable. 
Let us now increase $n$ further.
\item For $n=10^{-6}$, approximately the linear equations become 
\begin{eqnarray}
  \displaystyle \frac{d\delta X}{d \alpha}& =& -3.0014\, \delta X,      \label{sta2a} \\
 \displaystyle \frac{d\delta Y}{d \alpha} &=& -3.00125 \,\delta Y + 0.3946\, \delta Z ,    \label{sta2b} \\
 \displaystyle \frac{d\delta Z}{d \alpha}& = & -0.1962\, \delta  Y.     \label{sta2c}
\end{eqnarray}
Now, using (\ref{test_stab}) for analyzing stability, from (\ref{sta2a}), we get $\Omega=-3.0014$. 
Solving (\ref{sta2b}) and (\ref{sta2c}) give $\Omega= -2.9750, -0.0261$.
This shows that power-law anisotropic solutions are stable with this value of $n$.
We further increase $n$.
\item For $n=0.5 \times 10^{-6}$, we have following approximate linear equations 
\begin{eqnarray}
  \displaystyle \frac{d\delta X}{d \alpha}& =& -3.0030\, \delta X ,     \label{sta3a} \\
 \displaystyle \frac{d\delta Y}{d \alpha} &=& -3.0012 \,\delta Y + 0.5984\, \delta Z ,    \label{sta3b} \\
 \displaystyle \frac{d\delta Z}{d \alpha}& = & -0.2963\, \delta  Y.     \label{sta3c}
\end{eqnarray}
Using (\ref{test_stab}) to determining the stability, solution of (\ref{sta3a}) gives $\Omega=-3.003$. 
Solving (\ref{sta3b}) and (\ref{sta3c}) yield  $\Omega= -2.94, -0.060$.
Let us now increase $n$ by an order of magnitude higher.

\item  Then for $n=10^{-5}$, the approximate linear equations are 
\begin{eqnarray}
  \displaystyle \frac{d\delta X}{d \alpha}& =& -3.0127\, \delta X ,     \label{sta4a} \\
 \displaystyle \frac{d\delta Y}{d \alpha} &=& -3.0011 \,\delta Y + 1.2607\, \delta Z  ,   \label{sta4b} \\
 \displaystyle \frac{d\delta Z}{d \alpha}& = & -0.6287\, \delta  Y .    \label{sta4c}
\end{eqnarray}
Using (\ref{test_stab}) and solving (\ref{sta4a}) gives $\Omega=-3.0127$ 
while solution of (\ref{sta4b}) and (\ref{sta4c}) produce  $\Omega= -2.70, -0.29$  which
show that anisotropic solutions are stable for $n=10^{-5}$.
\end{itemize}
Starting from a small value i,e. $n=10^{-7}$, when $n$ is increased up to $n=10^{-5}$, we observe that 
anisotropic power-law solutions have stable fixed points (as $\Omega$ has negative eigenvalues). 
However, our study suggests anisotropic power-law solutions will possess stable fixed points 
with further increase in $n$ provided $\rho>>1,\lambda<<1, n<<1$ and $
(\lambda^2+2\lambda\rho) \left[1+\displaystyle\frac{2 n}{\lambda^4} (2+\lambda^2)\right]>4$ hold.
Since Gauss-Bonnet term is treated as higher order corrections of curvature in our analysis,
it is not desirable to increase $n$ to a large value which may drift away stable anisotropic 
fixed points from corresponding non-Gauss-Bonnet ($n=0$) fixed point solutions.
For example, in Section 4.1.2, it was found that with $\rho=40, \lambda=0.1$, the anisotropic point $X$ increases 
by an order of $10$ for $n=10^{-5}$ compared to $X$ evaluated at $n=0$.
\footnote{It is shown in \cite{Dadhich} that in the early Universe inflation may occur solely
due to the Gauss-Bonnet term such that Ricci scalar term can be completely ignored.}
This consideration led us to restrict our study till $n=10^{-5}$.\\

\section{Concluding remarks}
In the present work, we have studied anisotropic inflation in the backdrop of quadratic curvature corrections.
Anisotropic inflationary solutions are obtained in presence of Gauss-Bonnet correction term and 
a massless $U(1)$ gauge field whose 
kinetic part is coupled to the inflaton field.
The vector field plays a
significant role in generating anisotropic effects during slow-roll inflation. 
To bring about non-zero contribution of the Gauss-Bonnet term, it is non-minimally
coupled to the inflaton field.
In this scenario, we have showed that exact anisotropic power-law inflationary solutions can be constructed
when the inflaton potential, gauge kinetic function and the
Gauss-Bonnet coupling are exponential functions of the inflaton field.
In presence of Gauss-Bonnet corrections, we have derived a general relation which shows that anisotropy is
proportional to slow-roll parameters of the theory namely $\epsilon_H$ and $\delta_H$
as a consequence, anisotropy gets enhanced as compared to the non-Gauss-Bonnet case.
We have performed stability analysis and showed that obtained 
solutions are stable.\\
Thus our study shows that if Gauss-Bonnet corrections are taken into account, vector hair
may persist giving rise to enhanced anisotropic effects during slow-roll regime. 
This suggests that observational signatures of anisotropic Gauss-Bonnet inflation is worth studying. 
Furthermore our study hints that cosmic no-hair conjecture is required to be modified appropriately.
However, when the vector field does not have non-trivial contributions, 
anisotropy vanishes and isotropic inflation with Gauss-Bonnet corrections are realized.\\
Furthermore, it would be interesting to
study anisotropic inflationary solutions with Gauss-Bonnet gravity
in the context of most general scalar-tensor theory
particularly Horndeski theory so as to explore effects of higher order curvature corrections.

\section*{Acknowlegment}
I would like to thank Jiro Soda for illuminating suggestions and discussions at various stages of this work
and going through the manuscript.
I am thankful to Narayan Banerjee, Souvik Banerjee, Sugumi Kanno for helpful discissions
and to Claus Laemmerzahl in ZARM where this work was completed.

\bibliographystyle{JHEP}

 \end{document}